\documentclass[namedreferences]{solarphysics}
\usepackage[hyperref,optionalrh,solaromanenum]{spr-sola-addons} 

\usepackage{graphicx}                    
\usepackage{color}                       
\usepackage{breakurl}                         
\usepackage{footnote}
\usepackage{threeparttable}
\usepackage{amsmath} 
\usepackage{pdflscape}
\usepackage{ulem}
\usepackage{xspace}

\chardef\us=`\_

\begin{document}

\begin{article}

\begin{opening}

\title{Automatic detection of CMEs using synthetically-trained Mask R-CNN} 
%
\author[addressref={ff1,ff2},corref,email={francisco.iglesias@um.edu.ar}]{\fnm{}\lnm{*Francisco A. Iglesias}\orcid{0000-0003-1409-1145}}
\author[addressref={ff3,ff4},email={diego.lloveras@nasa.gov}]{\fnm{}\lnm{*Diego G. Lloveras}\orcid{0000-0003-1402-0398}}
\author[addressref={ff1},email={}]{\fnm{}\lnm{Florencia L. Cisterna}\orcid{}}
\author[addressref={ff1,ff2},email={}]{\fnm{}\lnm{Hebe Cremades}\orcid{}}
\author[addressref={ff1},email={}]{\fnm{}\lnm{Mariano Sanchez Toledo}\orcid{}}
\author[addressref={ff1,ff2},email={}]{\fnm{}\lnm{Fernando M. L\'opez}\orcid{}}
\author[addressref={ff1,ff2},email={}]{\fnm{}\lnm{Yasmin Machuca}\orcid{}}
\author[addressref={ff1,ff2},email={}]{\fnm{}\lnm{Franco Manini}\orcid{}}
\author[addressref={ff5},email={}]{\fnm{}\lnm{Andrés Asensio Ramos}\orcid{}}


%
\runningauthor{Iglesias \& Lloveras et. al.}
\runningtitle{Automatic detection of CMEs using Mask R-CNN}

\address[id={ff1}]{Grupo de Estudios en Heliofísica de Mendoza, Universidad de Mendoza, Boulogne Sur Mer 683, 5500 Mendoza, Argentina}
\address[id={ff2}]{Consejo Nacional de Investigaciones Científicas y Técnicas, Godoy Cruz 2290, C1425FQB CABA, Argentina}
\address[id={ff3}]{Heliophysics Science Division, NASA Goddard Space Flight Center, Greenbelt, MD 20771, USA}
\address[id={ff4}]{Physics and Astronomy Department, George Mason University, 4400 University Drive, Fairfax, VA 22030, USA}
\address[id={ff5}]{Instituto de Astrofísica de Canarias, 38205 La Laguna, Tenerife, Spain}
\begin{abstract}
Coronal mass ejections (CMEs) are a major driver of space weather. To assess CME geoeffectiveness, among other scientific goals, it is necessary to reliably identify and characterize their morphology and kinematics in coronagraph images. Current methods of CME identification are either subjected to human biases or perform a poor identification due to deficiencies in the automatic detection. In this approach, we have trained the deep convolutional neural model Mask R-CNN to automatically segment the outer envelope of one or multiple CMEs present in a single difference coronagraph image. The empirical training dataset is composed of $1.13\times10^5$ synthetic coronagraph images with known pixel-level CME segmentation masks. It is obtained by combining quiet (no CME) coronagraph observations, with synthetic white-light CMEs produced using the Graduated Cylindrical Shell geometric model and ray-tracing technique. To filter the different instances found by Mask R-CNN, we use the temporal consistency of mask properties such as the intersection over union ($IoU$). We found that our model-based trained Mask R-CNN infers segmentation masks that are smooth and topologically connected (without holes or isolated patches). While the inferred masks are not representative of the detailed outer envelope of complex CMEs, the neural model can better differentiate a CME from other radially moving background/foreground features, segment multiple simultaneous CMEs that are close to each other, and work with images from different instruments. This is accomplished without relying on kinematic information, i.e. only the included in the single input difference image. We obtain a median $IoU=0.98$ for $1.6\times10^{4}$ synthetic validation images, and $IoU=0.77$ when compared with two independent manual segmentations of 115 observations acquired by the COR2-A, COR2-B and LASCO C2 coronagraphs. The methodology presented in this work can be used with other CME models to produce more realistic synthetic brightness images while preserving desired morphological features, and obtain more robust and/or tailored segmentations. 
\end{abstract}

%
\end{opening}

\section{Introduction}
\label{sec:intro}
Coronal Mass Ejections (CMEs) involve the fast release of large amounts of mass and magnetic field from the solar corona into the interplanetary medium \citep[][]{Webb2012}. They produce significant perturbations in the solar wind and strongly influence space weather, with important detrimental impacts on various critical technologies \citep{zhang2018}. Due to the lack of routine magnetic field measurements of the solar corona \citep{judge2001}, among other reasons, no clear consensus has been reached regarding which are the predominant CME trigger and driving mechanisms \citep{Green2018}. This and other limitations result in our current inability to forecast when and where on the solar surface, the next CME is going to take place. Subsequently, it becomes critical to estimate the morphology and kinematics of CMEs as early as possible after their onset; in order to elaborate better predictions of their arrival time to Earth and geo-effectiveness.

CMEs manifest in white-light coronagraphs as bright radially-moving structures ejected from the Sun on timescales of minutes to several hours. Some emblematic space-based solar coronagraphs are the Large Angle and Spectrometric Coronagraph (LASCO, \citealt{Brueckner-etal1995}) onboard the Solar and Heliospheric Observatory (SoHO), which is located near the Lagrangian point L1; and the coronagraphs COR1 and COR2 part of the Sun-Earth Connection Coronal and Heliospheric Investigation (SECCHI, \citealt{Howard-etal2008}) package onboard the Solar-TErrestrial RElations Observatory (STEREO) twin spacecraft, that enables observing CMEs from  different perspectives in the ecliptic plane. Since their initial detection in the 1970s, there have been many efforts devoted to identify and characterize CMEs in white-light coronagraphic images, using various manual or automated techniques. We classify these techniques into three categories and summarize them below.

\textit{Manual}: Performed by an educated operator, such as in the LASCO CME Catalog\footnote{\hyperlink{http://cdaw.gsfc.nasa.gov/CME_list/index.html}{http://cdaw.gsfc.nasa.gov/CME\_list/index.html}} of the Coordinated Data Analysis Workshops (CDAW) Data Center \citep{Yashiro-etal2004, gopalswamy2009}. This approach can deal with complex scenarios that typically complicate image processing algorithms, such as events with fast-moving background material and/or multiple quasi-simultaneous events. On the other hand, manual inspection is subjective and thus not free of errors \citep[e.g.,][]{yashiro2008}. It is also time-consuming and impractical to apply to the large data volumes generated by present and proposed coronagraphs, such as the MOST \citep{gopalswamy2024} and Vigil missions \citep{palomba2022}.

\textit{Image processing}: This approach uses traditional image processing techniques, mostly based on pixel brightness spatio-temporal variations and fine-tuned filtering, performed in the height-time map associated to the event. This is done, for example, in the pioneering Computer-Aided CME Tracking method (CACTus; \citealt{Robbrecht-etal2009})\footnote{\hyperlink{http://sidc.oma.be/cactus/}{http://sidc.oma.be/cactus/}}, the Solar Eruptive Event Detection System (SEEDS; \citealt{Olmedo-etal2008})\footnote{\hyperlink{http://spaceweather.gmu.edu/seeds/}{http://spaceweather.gmu.edu/seeds/}}, the CORonal IMage Process (CORIMP; \citealt{Byrne-etal2012})\footnote{\hyperlink{http://alshamess.ifa.hawaii.edu/CORIMP/}{http://alshamess.ifa.hawaii.edu/CORIMP/}}, the Automatic Recognition of Transient Events and Marseille Inventory from Synoptic maps (ARTEMIS; \citealt{Boursier-etal2009})\footnote{\hyperlink{http://cesam.lam.fr/lascomission/ARTEMIS/}{http://cesam.lam.fr/lascomission/ARTEMIS/}}, and the CORonal SEgmentation Technique (CORSET, \citealt{goussies2010}) used at the multi-viewpoint STEREO catalog\footnote{\hyperlink{http://solar.jhuapl.edu/Data-Products/COR-CME-Catalog.php}{http://solar.jhuapl.edu/Data-Products/COR-CME-Catalog.php}} by \cite{vourlidas2017}. 
These image processing techniques have the advantage of being automatic or semi-automatic, with faster and easier implementation than manual inspection, and have resulted in useful CME catalogs (see the citations above). However, these algorithms generally fail in common cases, where the spatial and temporal brightness variation of CME and background/foreground moving material are difficult to distinguish from each other, see e.g., \cite{yashiro2008} and \cite{vourlidas2017}.

\begin{figure*}
  \centering
  \includegraphics[width=1\textwidth]{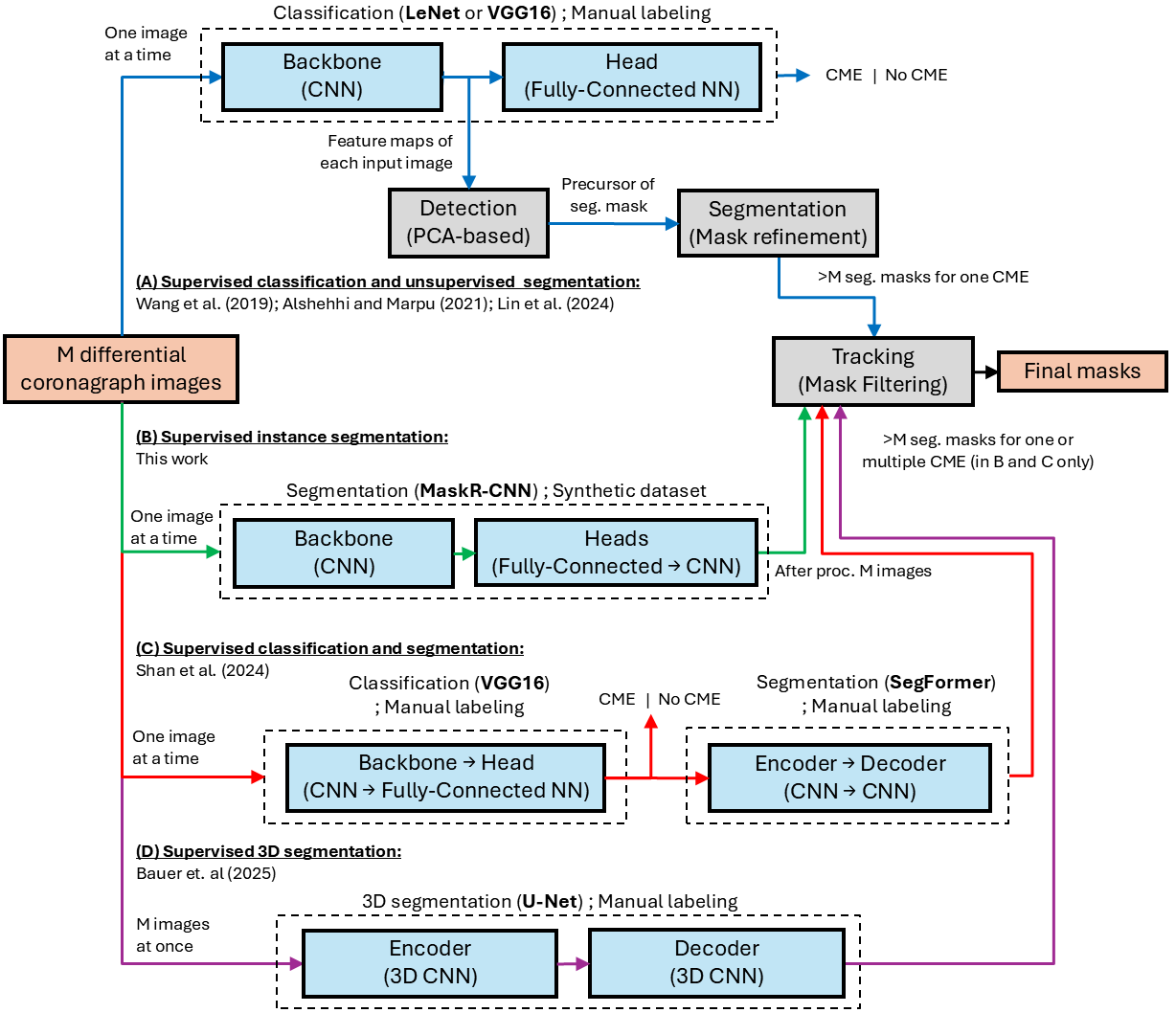}
  \caption{ML-based segmentation of CMEs. We summarize current approaches found in the literature, including supervised classification followed by unsupervised segmentation (A, blue arrows), supervised classification followed by supervised segmentation using manually labeled data (C, red arrows), and supervised 3D segmentation using manually labeled data (D, violet arrows). The supervised instance segmentation approach used in this work (B, green arrows) employs synthetic data instead of manual labeling. We sketch a simplified version of the data flow between blocks representing input/output (orange), different NNs (blue), and other processing algorithms (grey). The designation of each approach and the name of the neural model used are annotated in bold. See the text for extra details.}
  \label{fig:arch_comp}
\end{figure*}

\textit{Machine Learning}: Machine Learning (ML) has experienced enormous improvements during the last decade. Models involving Deep Convolutional Neural Networks (DCNNs) currently excel at solving machine-vision related tasks. This has led to various applications of ML to different topics in Solar Physics and Space weather, as reviewed by \cite{camporeale2019} and \cite{asensioramos2023}. The automatic detection of CMEs in coronagraph images using ML was pioneered by \citet{wang2019}, further explored by \cite{alshehhi2021} and very recently by \citet{lin2024}, \cite{shan2024} and \cite{bauer2025}. We summarize these and the present works in the diagram shown in Fig.~\ref{fig:arch_comp}. 

In the efforts grouped under approach A, a DCNN-based classifier is first trained to label images as CME or non-CME, typically using a backbone plus a fully connected head. \citet{wang2019} and \citet{lin2024} employ as backbone LeNet trained on manually classified LASCO CDAW images, whereas \cite{alshehhi2021} use the VGG16 model \citep{simonyan2014} already trained for classification with the Imagenet database \citep{deng2009}, which does not contain CME images. The deep feature maps from the classifier are then repurposed to derive a coarse mask precursor in an unsupervised manner: for each image of the event a feature map is extracted, and Principal Components Analysis (PCA) is applied to exploit spatio–temporal correlations. \citet{wang2019} and \citet{lin2024} implement the PCA-based image co-location method of \cite{wei2019}, while \cite{alshehhi2021} combine PCA with K-means clustering. This precursor is converted into a binary mask via post-processing. \citet{wang2019} use a graph-cut formulation that also accounts for temporal expansion; \cite{alshehhi2021} apply morphological erosion and dilation to smooth and improve connectivity; and \citet{lin2024} threshold using Otsu’s method \citep{otsu1979} followed by the same morphological operators. Finally, a tracking stage enforces temporal consistency and removes spurious regions by leveraging expected CME behavior (e.g., outward motion, small central position angle changes). \citet{wang2019} and \cite{alshehhi2021} discard all pixels at position angles whose maximum height is less than half the image field of view, and reject any CME candidate that does not move outward in at least two consecutive frames or does not extend beyond the LASCO C2 field of view \citep[following][]{Olmedo-etal2008}. On the other hand, \citet{lin2024} adopts a more general similarity metric, based on a weighted sum of differences in geometric centers, areas, and shape descriptors (the latter following \citealt{chaumette2004}).

Two alternatives to approach A are using fully unsupervised learning, e.g., with an auto-encoder architecture \citep{xia2017}, not explored so far; or producing a suitable training dataset for supervised segmentation, i.e. including a known true mask for each image. Supervised segmentation has the advantage that the error between inferred ($\bar{M}$) and true ($M$) masks can be quantified using common metrics, such as Intersection over Union ($IoU$) or the Relative Symmetric Difference ($RSD$), see e.g., \cite{reinke2021}:
\begin{align}
    \label{eq:iou}
    IoU = \frac{\bar{M} \cap M}{\bar{M} \cup M},\, 
    RSD = \frac{\bar{M} \cup M - \bar{M} \cap M}{M}.
\end{align}

Training data can be generated via manual pixel-level labeling done by a human operator, or synthetically, using a coronal and/or CME model. The former was explored recently by \cite{shan2024}, approach C in Fig.~\ref{fig:arch_comp}, who manually labeled 375 LASCO C2 images to train the SegFromer model \citep{xie2021}; and by \citet{bauer2025}, approach D in Fig.~\ref{fig:arch_comp}, who produced $\approx$\,8400 labels for images of the Heliospheric Imagers \citep[HI,][]{eyles2009} onboard STEREO and trained the 3D U-Net model \citep{ronneberger2015}. \cite{shan2024} introduce an improved multi-CME tracking algorithm and report $IoU=0.83$, which is a significant improvement with respect to the $0.49$ of their previous work in \citet{wang2019}. \textbf{\cite{shan2024} have also released a 2D segmentation catalog, and used this technique to produce also a 3D CME catalog, see the reference for further details.} On the other hand, \cite{bauer2025} uses a beneficial 3D DCNN that allows processing multiple images of a single event at once and reports $IoU=0.39$. We note that the latter uses data from HI, which observes CMEs much farther from the Sun (up to 330 R$_{\odot}$), and thus the images are typically noisier, and the fainter CMEs are considerably more difficult to differentiate from moving background/foreground material.

The works presented in Fig.~\ref{fig:arch_comp}, and briefly described above, are important contributions to the advancement of automatic CME segmentation and characterization. Their results show that ML-based approaches are generally better than image processing algorithms at detecting weaker CMEs in earlier stages, smaller changes in the identified CME region, CMEs in more complex scenarios such as those containing fast-moving background/foreground material, and CMEs with smaller angular widths. \cite{alshehhi2021} also shows that their method is closer to the results of the manual CDAW catalog than to those of the CACTus and CORIMP algorithms for 2000 random CMEs.

Besides these promising results, there are still limitations. In approach A, the final segmentation mask properties are strongly influenced by the unsupervised algorithms that derives them from the neural feature maps. The same occurs with the algorithms used for mask filtering in the tracking step of approaches C and D. These algorithms are based on spatio-temporal correlation plus filtering and thresholding, making it difficult to generalize to all CME cases and stages of the solar cycle, and resulting in common errors where nearby moving background/foreground material is segmented as part of the CME bulk. On the other hand, the subjective manual labeling of \cite{shan2024} and \cite{bauer2025} also presents disadvantages, some already pointed out by the authors. The labeling process is very time-consuming, and different operators can produce different masks for slow and/or faint CMEs that do not present a strong contrast with respect to the background, among other subjective effects. In addition, there is no precise definition of the CME boundary, which harms the usage of quantitative error metrics to compare the performance of different methods in a systematic manner.

In this work, denoted approach B in Fig.~\ref{fig:arch_comp}, we explore the usage of synthetic data for supervised training. This has been successfully done in other Solar Physics applications, such as spectropolarimetric Stokes inversions \citep{milic2020, gafeira2021} and photospheric velocity estimations \citep{asensioramos2017}, among others. Our approach implies generating \textbf{a dataset} composed of synthetic CME coronagraph images with known pixel-level segmentation masks, using it to train a commercial-off-the-shelf \textbf{Deep Neural Network (DNN)} to perform CME segmentation, and using the trained model to infer CME masks from real coronagraph images. Even when this approach presents limitations related to the quality of the empirical synthetic data, see Sect.~\ref{sec:data_and_methods}, it also helps tackling some of the drawbacks discussed above. Particularly those related to incorporating high-level morphological properties of the CME mask to help better differentiate it from other moving material, and thus rely less on the mask filtering step. We demonstrate the potential of this approach using synthetic data based on a simple morphological CME model, namely the Graduated Cylindrical Shell (GCS, see Sect.~\ref{sec:data_and_methods}), in combination with ray-tracing. However, the same approach can be used with any other model that allows generating, potentially more realistic, synthetic CME brightness images and their corresponding masks. 

The rest of the paper is organized as follows. Sect.~\ref{sec:data_and_methods} details the generation of the synthetic training data set (Sect.~\ref{sec:syn_dataset}), the architecture and training of the DNN used for CME segmentation (Sect.~\ref{sec:model}). The results are presented and discussed in Sect.~\ref{sec:results}, including the performance on synthetic (Sect.~\ref{sec:results_syn}) and manually segmented CME observations (Sect.~\ref{sec:result_real_obs}). The conclusions and outlook are given in Sect.~\ref{sec:conclusions}.

\section{Data and Methods}
\label{sec:data_and_methods}

\begin{figure*}
  \centering
  \includegraphics[width=0.8\textwidth]{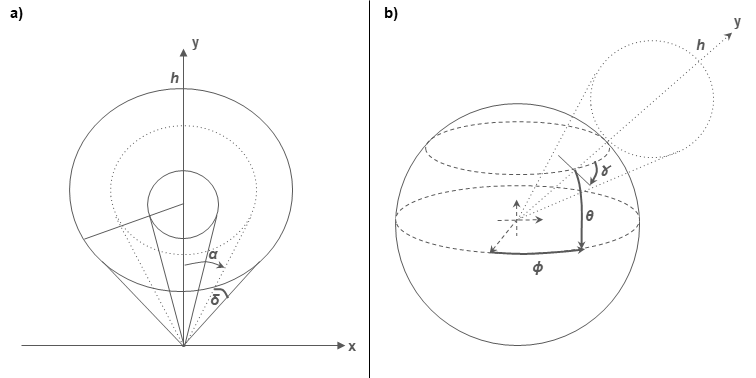}
  \caption{The GCS geometric model, including its face-on view (\textit{panel a}) and a 3D perspective showing the source region on the solar surface (\textit{panel b}). The 6 model parameters are the source region Stonyhurst longitude ($\phi$) and heliographic latitude ($\theta$), tilt angle with respect to the local parallel ($\gamma$), height of the apex ($h$), aspect ratio ($\kappa$), and half angular separation of the legs ($\alpha$). Adapted from \cite{Thernisien-etal2009}.}
  \label{fig:gcs_model}
\end{figure*}

The synthetic images, generated for training purposes, are formed by adding a synthetic CME differential brightness image to a differential background image taken from archive coronagraph observations. The synthetic CME images are based on the empirical Graduated Cylindrical Shell model \citep[GCS,][]{Thernisien-etal2006}, which defines a 3D geometric shape to describe the flux rope structure of CMEs. The GCS has a ``croissant-like" shape, with a circular cross section and two conical legs anchored to the center of the Sun (see Fig.~\ref{fig:gcs_model}). The 3D shape and its position are completely defined by the six following parameters:

\begin{itemize}
    \item $\phi, \theta$: Stonyhurst longitude and Heliographic latitude of the intersection point between the GCS main axis and the solar surface, respectively.
    \item $\gamma$: Tilt angle with respect to the local parallel.
    \item $h$: Height of the apex.
    \item $\kappa = \sin{\delta}$: Aspect ratio, where $\delta$ is the half angular width of the legs.
    \item $\alpha$: Half the angular separation of the legs.
\end{itemize}

The GCS is unable to model in detail the outer shell shape of complex CMEs. However, with only six degrees of freedom, the model defines a topologically connected shape that retains important basic properties attributed to CMEs, such as a self-similar bulk expansion. The GCS is widely used to reconstruct the 3D structure of the CME outer shell, an inherently ill-posed problem due to the scarce coronagraphic (2D) data available for any given event, see e.g., \cite{Mierla-etal2009,Antunes-etal2009,Jackson-etal2010, Mierla-etal2010, Thernisien-etal2011, anand2011,Joshi-Srivastava2011, Feng-etal2013, Cremades2020, Sieyra2020, verbeke2022}. The 3D reconstruction with GCS is done in a manual and subjective way, by visually matching its 2D projections to coronagraphic images acquired quasi-simultaneously from 2 or 3 different vantage points, see e.g., \cite{verbeke2022}. 

To isolate the CME signal from the background coronal emission, running-difference and base-difference techniques are usually applied to white-light coronagraphic images \citep{Vourlidas-etal2000, Vourlidas-etal2010}. However, there are dynamic background coronal structures (e.g., streamers) that are not completely removed by difference techniques, particularly at lower heights close to the telescope's central occulter. To make the synthetic data more realistic, we include these coronal structures in our synthetic training dataset by using as background actual running-difference coronagraphic images, which are manually checked to contain no CMEs (see Sect.~\ref{sec:syn_dataset}). We employ data obtained by SoHO/LASCO C2, and SECCHI/COR2-A and B \textbf{coronagraphs}. Both instruments detect the Thomson scattering of photospheric light by free electrons in the solar corona \citep{Billings1966}, \textbf{with an effective field of view (FOV) of 2.2}\,--\,6\,R$_{\odot}$ and 2.5\,--\,15\,R$_{\odot}$ for C2 and COR2, respectively. 

\subsection{Synthetic CME dataset}
\label{sec:syn_dataset}
\begin{figure*}
  \centering
  \includegraphics[width=1\textwidth]{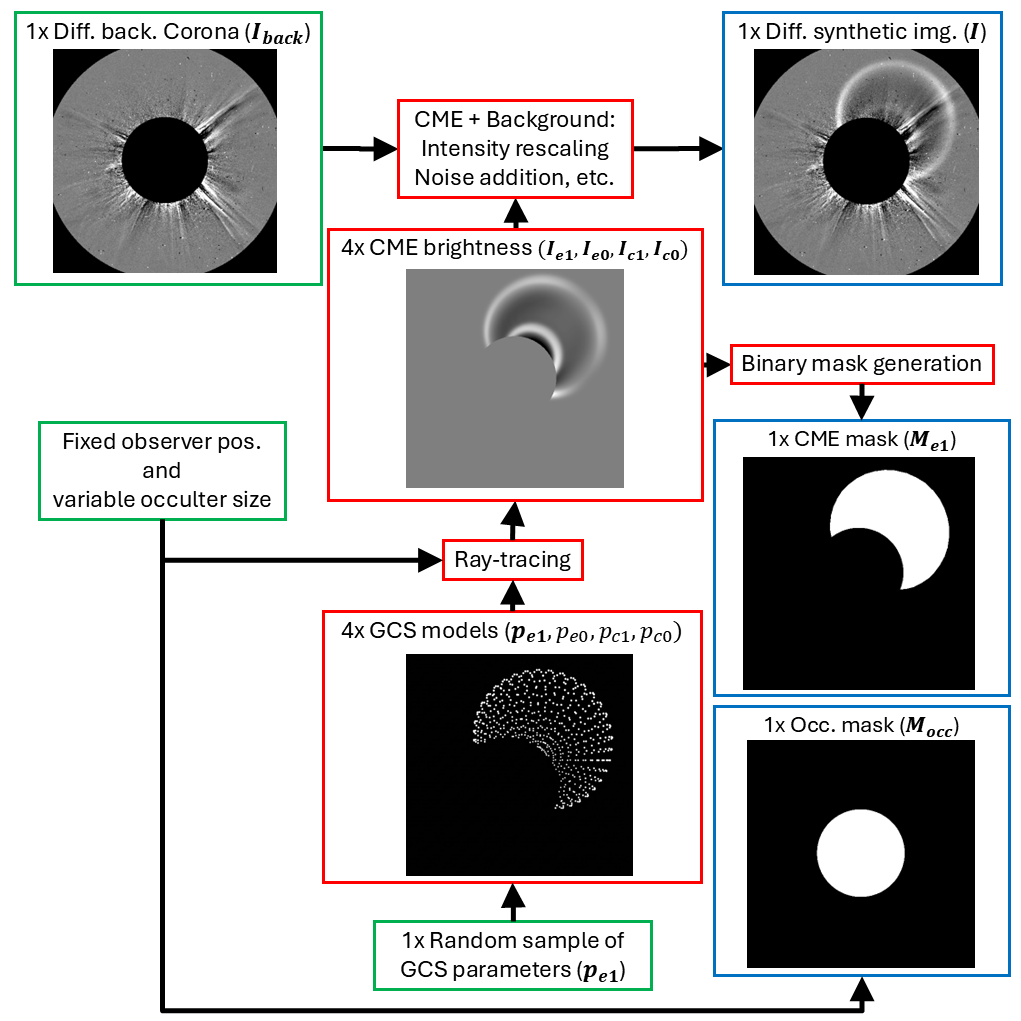}
  \caption{\textbf{Generation of the empirical synthetic dataset. Each dataset element is composed of a differential synthetic coronograph image and its corresponding CME outer shell and occulter binary segmentation masks (blue boxes). These are produced using a real observed coronal background (with no CME) and synthetic brightness images obtained from a ray-tracing simulation based on a random GCS model (green boxes). The red boxes represent intermediate steps. Note that we show only one of the four GCS models for simplicity. We deliberately selected a case that highlights the main drawback of this approach, namely, the lack of correlation between CME and background structures. See the text for extra details.}}
  \label{fig:syn_data}
\end{figure*}

Each element of the dataset used for training the DNN is composed of a synthetic coronagraphic image and its corresponding occulter and CME outer shell binary segmentation masks. \textbf{This synthetic image represents a single running difference snapshot of a CME event, i.e. we produce only two time instants per event}. The procedure used to obtain each element is outlined in Fig.~\ref{fig:syn_data} and detailed below:
\begin{enumerate}
    \item \textit{Selection of background coronal image:} We manually select running-difference coronagraphic images that do not contain CMEs to use as background to our synthetic CMEs. In the first step, one background image ($I_{back}$) is randomly selected from the backgrounds pool. Our backgrounds pool has 213 images: $45$, $39$ and $129$ from LASCO C2, COR2-A, and COR2-B, respectively, covering the minimum and rising phase of solar cycle 24 from the year 2007 to 2011. The standard Level 1 processing, made available in the SolarSoft library\footnote{\url{https://www.lmsal.com/solarsoft/}} by the respective instrument teams, is applied to the observations from the three coronagraphs. Subsequently, images are saved as binned, running-differences, with a size of $512\times512\,\text{pixel}^2$.
    \item \textit{Selection of GCS parameters:}  We defined the following subspace of the GCS parameters based on 
    CME statistical properties \citep[see e.g.,][]{Cremades-Bothmer2004, vourlidas2017, Kay-etal2023}:
    \begin{equation}
        \mathbf{P}=\{(-180,180),(-70,70),(-90,90),(2.0,20),(0.2,0.6), (5,65)\}
    \end{equation}
    This subspace is uniformly sampled using the Latin Hypercube method \citep{mcKay1979}, i.e., dividing the ranges in $\mathbf{P}$ by the desired size of the training dataset. Then, a point:
    \begin{equation}
        \mathbf{p_{e1}}=(\phi,\theta,\gamma,h,\kappa,\alpha),
    \end{equation}    
    is obtained by selecting a uniform random value from each GCS parameter in $\mathbf{P}$. 
     From $\mathbf{p_{e1}}$, we derive three more sets of parameters to obtain two pairs, namely $\mathbf{p_{e0}, p_{e1}}$ and $\mathbf{p_{c0}, p_{c1}}$. Each pair is used to produce the differential image of the CME envelope and its core, respectively. The first pair models the outer envelope with parameters $\mathbf{p_{e1}}$ and a smaller envelope corresponding to a previous time with the parameters:
    \begin{equation}
         \mathbf{p_{e0}}=(F_0\phi,F_1\theta,F_2\gamma_0,h-F_3,F_4\kappa,F_5\alpha),
    \end{equation}
    where $F_0$, $F_1$, and $F_2$ are uniform random numbers in the $(0.95,1.05)$ range, simulating the possible deflection and rotation of the CME, see e.g., \cite{Sieyra2020}. $F_3$ is the CME displacement computed using a random velocity and assuming a cadence of 15 min (typical for the instruments used). The random velocity is derived from a log normal random distribution with mean and standard deviation of $\log(433)$ and $0.5267$, respectively (\citealt{Yurchyshyn2005}), and forced to be in the $(200, 1800)$ km\,s$^{-1}$ range (\citealt{Vourlidas-etal2010}). $F_4$ and $F_5$ are uniform random numbers in the $(0.9,1)$ range, simulating the non-self similar expansion of the CME, see e.g., \cite{Cremades2020}. 
    
    The second pair of GCS parameters is used to simulate the CME bright core, a conspicuous feature typically present in CMEs and believed to generally lie at the bottom of a magnetic flux rope \citep{Vourlidas-etal2013}. Their values are:
    \begin{equation}
         \mathbf{p_{c1}}=(\phi,\theta,\gamma,F_6h,F_7\kappa,\alpha)
    \end{equation}
    and
    \begin{equation}
         \mathbf{p_{c0}}=(F_0\phi,F_1\theta,F_2\gamma,F_6h-F_3,F_4F_7\kappa,F_5\alpha),
    \end{equation}
    where $F_6$ is a uniform random number in the range $(0.55,0.85)$ or $(0.45,0.75)$ if $h_0$ is below or above 3~R$_{\odot}$, respectively, which controls the bright core height \citep{anand2011, sarkar2019}. $F_7$ is a uniform random number in the $(0.2,0.4)$ range, which defines the bright core smaller aspect ratio.
    \item \textit{Generation of synthetic image}: We run the ray-tracing implementation by \cite{Thernisien-etal2009} in the four GCS models with parameters $\mathbf{p_{e1}}$, $\mathbf{p_{e0}}$, $\mathbf{p_{c1}}$ and $\mathbf{p_{c0}}$ to obtain four $512\times512\,\text{pixel}^2$ corresponding synthetic images: $I_{e1}$, $I_{e0}$, $I_{c1}$, and $I_{c0}$. These images model the total brightness scattered by the CME outer envelope and bright core as observed from a given observer position, defined by [Carrington longitude, heliographic latitude, tilt with respect to the Solar North]. We use the fixed positions [300.082, 1.955, 0.000], [32.894, 7.051, 0.000], or [170.089,  -6.587,   0.000]~deg when using COR2-A, COR2-B or LASCO C2 background coronal images, respectively. The differential CME image is then obtained from:
    \begin{equation}
        I_{cme} = F_{10}[I_{e1}-F_8I_{e0}+F_9(I_{c1}-F_8I_{c0})],
    \label{eq:Icme_def}
    \end{equation}
    where $F_8=(1-F_3/h)^{-3}$ is a scaling factor based on the expected density reduction of an expanding CME given in \cite{howard2018}; $F_9$ is an ad-hoc uniform random number in the $(0,2)$ range that controls the relative levels of the differential CME bright core and outer envelope; and $F_{10}$ is an ad-hoc random scaling map that varies smoothly in the $(0.1,1)$ range across the image. The latter is added to break the homogeneity of the brightness images produced by the ray-trace simulation of the simple and smooth GCS shape (which is far from the non-uniform brightness observed in real coronagraphic images). 
    
    The image $I_{cme}$ is normalized to the $(0,1)$ range to obtain $\hat{I}_{cme}$, and assemble the final output image as follows:

    \begin{equation}
        I = F_{11}\sigma(I_{back})\hat{I}_{cme} + I_{back} + I_{occ} + I_{noise},
        \label{eq:I_tot}
    \end{equation}   
    where $F_{11}$ is a uniform random number in the $(2,7)$ range and $\sigma(I_{back})$ is the standard deviation of the background image. The product $F_{11}\sigma(I_{back})$ is used to ensure the CME image is visible over differential backgrounds of varying contrasts, basically because the CME density assumed during the ray-tracing simulations is a constant number uncorrelated to the background coronal density that produced $I_{back}$. The term $I_{occ}$ is an artificial occulter of radius equal to the one present in $I_{back}$ plus a random variation in the $(1,1.1)$ range, added to promote a more instrument independent response of the DNN. Finally, the term $I_{noise}$ represents additive Poisson noise computed using $\hat{I}_{cme}$. 
    
    \item \textit{Generation of binary segmentation masks:} We derive an outer envelope binary segmentation mask ($M_{e1}$) from $I_{e1}$, with value equal to one for all pixels in the interior of the geometrical border defined by the projection of $\mathbf{p_{e1}}$ and outside of the occulter, see Fig. \ref{fig:syn_data}. We also obtain the central occulter binary mask $M_{occ}$ from $I_{occ}$. Given that we do an independent random sampling for each of the six GCS parameters in the subspace $\mathbf{P}$, there can be cases where the projected CME ($I_{CME}$) lies only in the central area of the image occupied by the occulter, e.g., a small halo CME. These cases are eliminated from the dataset because $M_{e1}$ is null. We also filter out any event that has $M_{e1}$ fragmented by the occulter into two disconnected areas, e.g., in a halo CME of projected size similar to the occulter diameter. 
    \textbf{Note that we train Mask R-CNN to segment only the CME outer envelope, thus no mask is derived for the CME bright core ($\mathbf{p_{c1}}$ and $ \mathbf{p_{c1}}$). This feature is added to the synthetic image to make the NN segmentation more robust against the strong intensity contrast typically observed in the interior of the CME envelope.}
\end{enumerate}

The above-described procedure to generate synthetic images is based on the simple GCS model and includes various empirical parameters, thus presents the following main disadvantages:
\begin{itemize}
    \item There is no correlation between the CME morphology and the background coronal structures, which is unrealistic. \textbf{We deliberately chose an example that highlight this in Fig.~\ref{fig:syn_data}}. For example, there are frequent observations of deflected streamers \citep[e.g.,][]{Vourlidas-etal2003, Vrsnak-etal2006, Tripathi-Raouafi2007} or streamer blowouts \citep{Sheeley1982, Vourlidas-Webb2018} associated to CMEs, among others. 
    \item We include an unrealistic random component of the CME brightness spatial variation that is not correlated with its morphology via $F_{10}$, see Eq.~\ref{eq:Icme_def}.
    \item The GCS gives a smooth geometric shape that is not representative of many real CMEs outer shells or bright cores, see e.g., \citet{Pluta2019, verbeke2022}.    
\end{itemize}
And the following advantages:
\begin{itemize}
    \item It can be used to efficiently generate as many synthetic images as necessary, to better investigate the training of large DNNs.
    \item By using a real observed coronal background ($I_{back}$), the dataset includes some of the complexity of the solar corona in different stages of the solar cycle. 
    \item The desired properties of the output mask can be controlled based on the CME model used. For example, the GCS produces masks that are fully connected, and thus it is expected that the DNN learns to produce similar masks. This reduces the common issue of mixing CME and background fast moving material during segmentation of real observations, see Sect.~\ref{sec:result_real_obs}.  
    \item Our training dataset is public to help quantitatively compare the performance of different CME segmentation algorithms\footnote{See \url{https://sites.google.com/um.edu.ar/gehme/science/cme-segmentation}}. 
\end{itemize}
We note that the above-named disadvantages are mostly related to the simplicity of the morphological and CME density models used. However, the approach presented in this work remains valid if a more accurate coronal and/or CME model is used to generate the synthetic images (see Sect.~\ref{sec:conclusions}).

\subsection{Neural CMEs segmentation}
\label{sec:model}
\subsubsection{Architecture and supervised training}
\begin{figure*}
  \centering
  \includegraphics[width=0.85\textwidth, trim=0.3cm 0cm 0cm 0cm, clip]{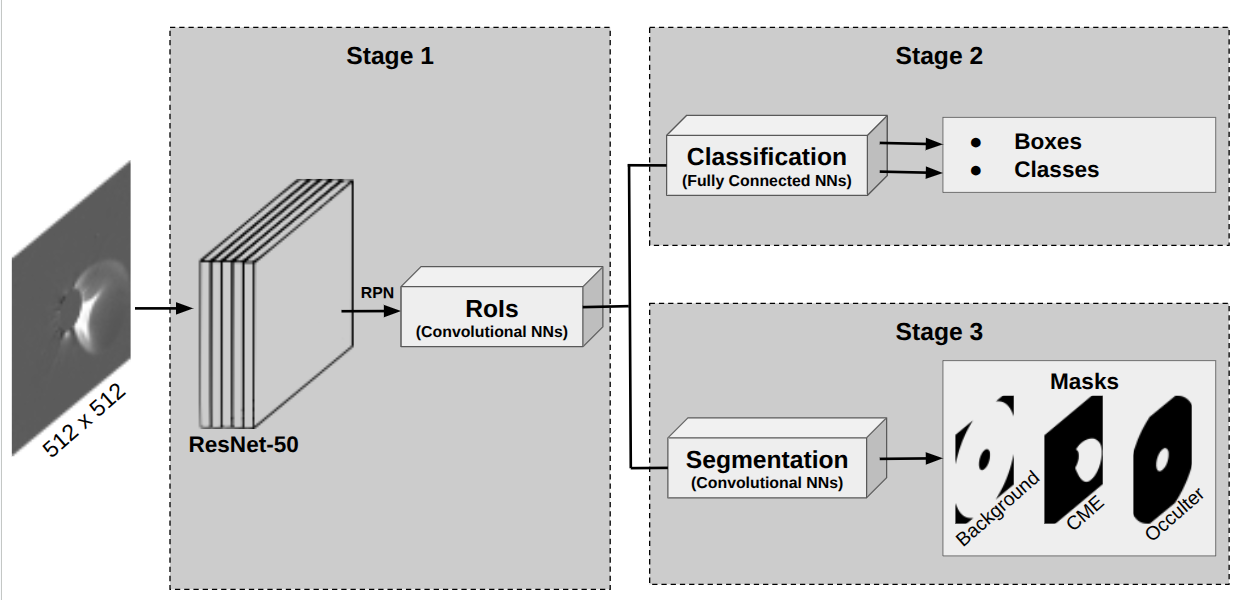}
  \caption{Architecture of the Mask R-CNN model \citep{he2017} used for instance segmentation of CMEs in coronagraphic images. The backbone is a ResNet-50 with initial weights trained on the COCO dataset. We use the object classes: Background, CME and Occulter.}
  \label{fig:maskrcnn}
\end{figure*}

We employ the PyTorch implementation of the Mask Region-based Convolutional Neural Network\footnote{Downloadable at   \url{https://docs.pytorch.org/vision/main/models/mask\_rcnn.html}} \citep[Mask R-CNN,][]{he2017}, which is a very successful DNN used for identifying and segmenting objects within images across various domains, including astronomy \citep{burke2019,farias2020,pearce2025}. Mask R-CNN consists of three stages, see Fig.~\ref{fig:maskrcnn}. The first two stages are identical to those used in the precursor Faster R-CNN detector \citep{ren2015}. The first stage, named Region Proposal Network, proposes candidate object bounding boxes. The second stage extracts features from each possible box and performs classification. The core of Mask R-CNN lies in the third stage, parallel to the second, that provides a mask for each region of interest. As backbone, we use the Residual Network (ResNet, \citealt{kaiming2015}), which addresses the vanishing gradient problem when training very deep CNN by using residual connections. We employ a 50-layers ResNet (Resnet-50) that has initial weights obtained by training it for image classification with the Common Objects in Context \citep[COCO,][]{tsung-yi2014} dataset. 

In our setup, the Mask R-CNN input is a single $512\times512\,\text{pixel}^{2}$ differential coronagraphic image, while the outputs are the segmentation masks, class scores and bounding boxes the model found in the input. We apply a $3\times3\,\text{pixel}^{2}$ median filter to reduce noise and then normalize the input images such that the range $[{\mu_I}-2{\sigma_I}, {\mu_I}+2{\sigma_I}]$ is mapped to $[0,1]$. All pixels outside of this range are clipped. We define three object classes, two task-specific: \textit{CME} and \textit{Occulter}, and the \textit{Background} required by the model. Each output mask is an array that gives for each pixel within the bounding box the inferred likelihood of the pixel belonging to the object (given by the class score). By thresholding each output mask (see Sec~\ref{sec:mask_gen}), a corresponding binary segmentation mask can be derived ($\bar{M}^i_{e0}$). We use the upper index $i$ to emphasize that Mask R-CNN is able to do instance segmentation, therefore multiple masks can be output for a single input image (see Sect.~\ref{sec:instance_selec}).

\begin{figure*}
  \centering
  \includegraphics[width=1\textwidth]{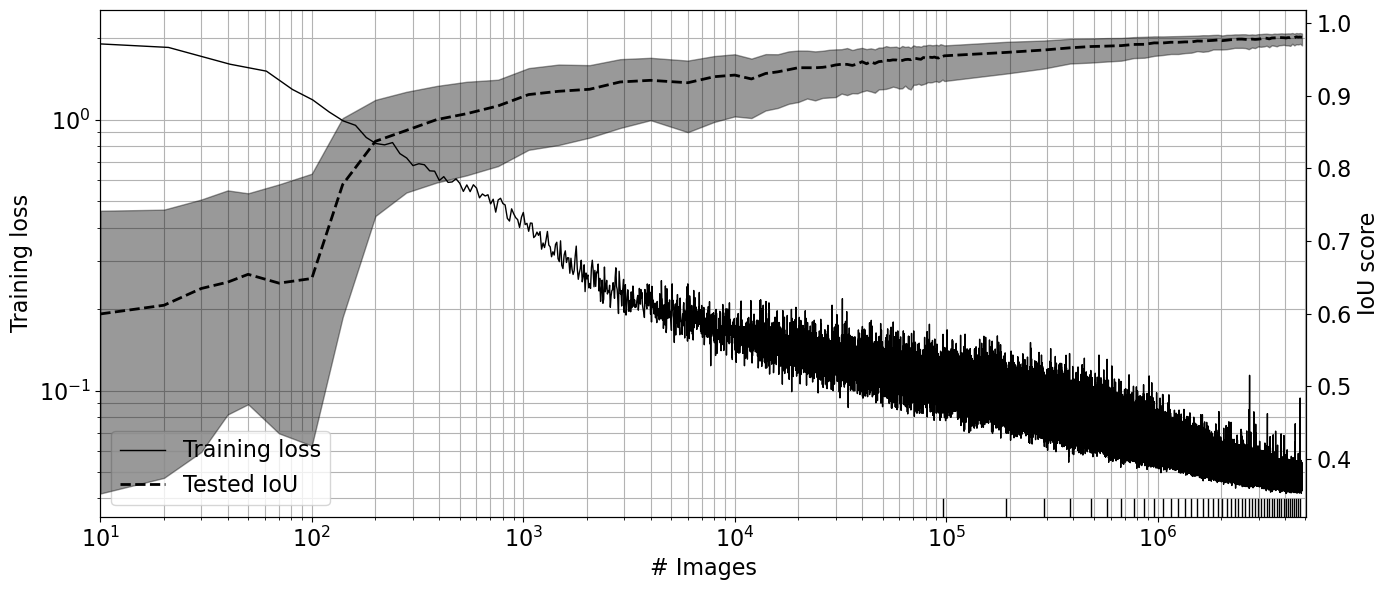}
  \caption{Mask R-CNN model training on synthetic coronagraphic images. We show the training loss (solid line) and mean validation IoU (dashed line, refer to the right vertical axis) as a function of the number of training images. The gray band represents the 25th-75th percentile of the $IoU$, which is computed for the validation set. Black vertical segments on the horizontal axis mark each training epoch.}
  \label{fig:iou_vs_epoch}
\end{figure*}

We generate a synthetic dataset of $1.13\times10^5$ images using the methodology described in Sect.~\ref{sec:syn_dataset}. This dataset is split into $85\%$ for training and $15\%$ for validation. We train Mask R-CNN using a batch size of 20 images for {50} epochs, reaching a loss function of $\approx0.5$, see Fig.~\ref{fig:iou_vs_epoch} and \cite{he2017} for the exact definition of the loss. Given that in our synthetic dataset the true CME outer shell mask is known, we can quantify the error of each inferred mask by computing $IoU$ between $M_{e0}$ and each $\bar{M}^i_{e0}$, see Eq.~\ref{eq:iou}.

\subsubsection{Binary Segmentation Mask}
\label{sec:mask_gen}
\begin{figure*}
  \centering
  \includegraphics[width=0.85\textwidth]{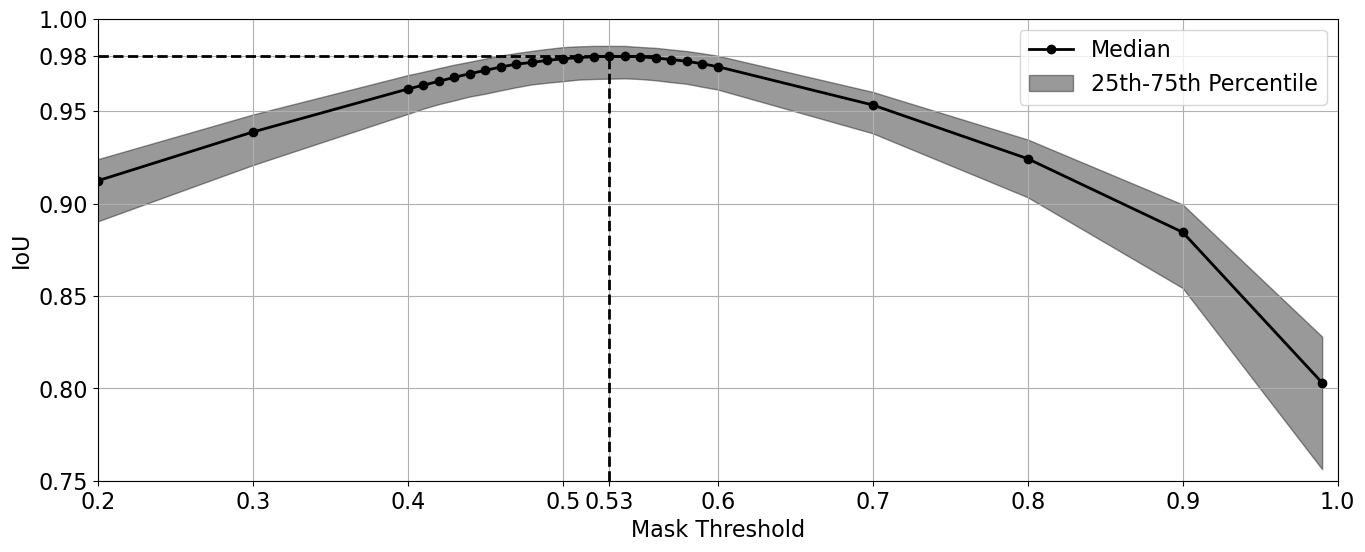}
  \caption{$IoU$ versus mask threshold. The solid line and gray area indicate the median and 25th-75th percentiles of the $IoU$ computed for $10^3$ random images of the synthetic validation dataset. The dashed vertical and horizontal lines indicate the maximum.}
  \label{fig:iou_vs_mt}
\end{figure*}

We derive the final binary segmentation masks by defining a threshold such that all pixels in the Mask R-CNN output masks with values above the threshold are considered belonging to the object. In many applications, this mask threshold is typically in the $0.60-0.85$ range \citep[e.g.,][]{lopezcano2024}. We found that such large values lead to underestimations of the commonly diffuse and faint CME outer shell. We select a task-specific value via a simple hyperparameter search. For different values of the mask threshold, we compute the median $IoU$ for $10^3$ random synthetic validation images. We then selected the mask threshold value that maximizes such a validation $IoU$, see Fig.~\ref{fig:iou_vs_mt}. We obtain an optimal mask threshold value of 0.53, which is used for the rest of this work.

\subsubsection{Instance selection via CME tracking}
\label{sec:instance_selec}

Given that Mask R-CNN can perform instance segmentation, multiple CME masks can be produced for a single input image, including overlapping objects. This has the advantage that cases with multiple simultaneous CMEs can be identified, and the disadvantage of possible false positive detections. In other applications, this is typically addressed by keeping only outputs with class scores larger than a predefined threshold, after possibly eliminating overlapping objects using $IoU$. However, we found that this approach cannot be consistently used to select the best CME mask, particularly in cases of very faint and diffuse borders, nor to differentiate a false positive detection from the correct detections in images of multiple CMEs events. Another approach, applicable when a time series of a CME event is available, is to use the expected temporal behavior of CME mask properties to select the correct instance. This is the CME tracking step included in all methods shown in Fig.~\ref{fig:arch_comp}. The main difference is that the other techniques use this step to eliminate undesired portions of the mask(s), altering their final shape. On the contrary, we use CME tracking only to select the best of the multiple masks found in each input image. The procedure, similar to that given in \cite{shan2024} but simpler, is outlined below.

We start with a time series of $M$ differential images of a CME event. For each image, our trained Mask R-CNN infers (possibly) multiple CME binary masks. By selecting one mask per image, we obtain a specific temporal sequence of masks. Firstly, we compute all the possible sequences in the time series, the goal is to find the sequence that best resembles a radially propagating CME. Secondly, we compute the apex (percentile 98 of the radial distance distribution of the mask pixels) of all the masks in each sequence. Each sequence that includes two temporally consecutive masks with apex difference smaller than that corresponding to a radial growth of 200 km\,s$^{-1}$ (lowest yearly average radial speed, proxy value from \citealt{vourlidas2017}) is eliminated. Thirdly, we compute for each sequence a total error, which quantifies the similarity of the masks. The error between the consecutive masks number $n$ and $n+1$ in a sequence is:
\begin{equation}
          \Delta_{n} = \omega_{CPA} \cdot \Delta_{CPA} + \omega_{AW} \cdot \Delta_{AW} + \omega_{IoU} \cdot \Delta_{IoU},
\end{equation}
where $\Delta_{CPA}$ and $\Delta_{AW}$ are the differences of the masks Central Position Angles (CPAs) and Angular Widths (AWs), respectively. The CPA and AW of each mask are defined as the median, and the difference between percentile 95 and 5, of the position angles distribution, respectively. $\Delta_{IoU}$ is the inverse of the IoU between the masks, and the $\omega$ are normalization constants used to keep each term of the sum in the 0-1 range. Finally, the mask sequence that minimizes the total error, $\sum_{n=1}^{M}\Delta_n$, is selected.

\section{Results and discussion}
\label{sec:results}

\subsection{Performance on synthetic validation data}
\label{sec:results_syn}
\begin{figure*}
  \centering
  \includegraphics[width=0.6\textwidth]{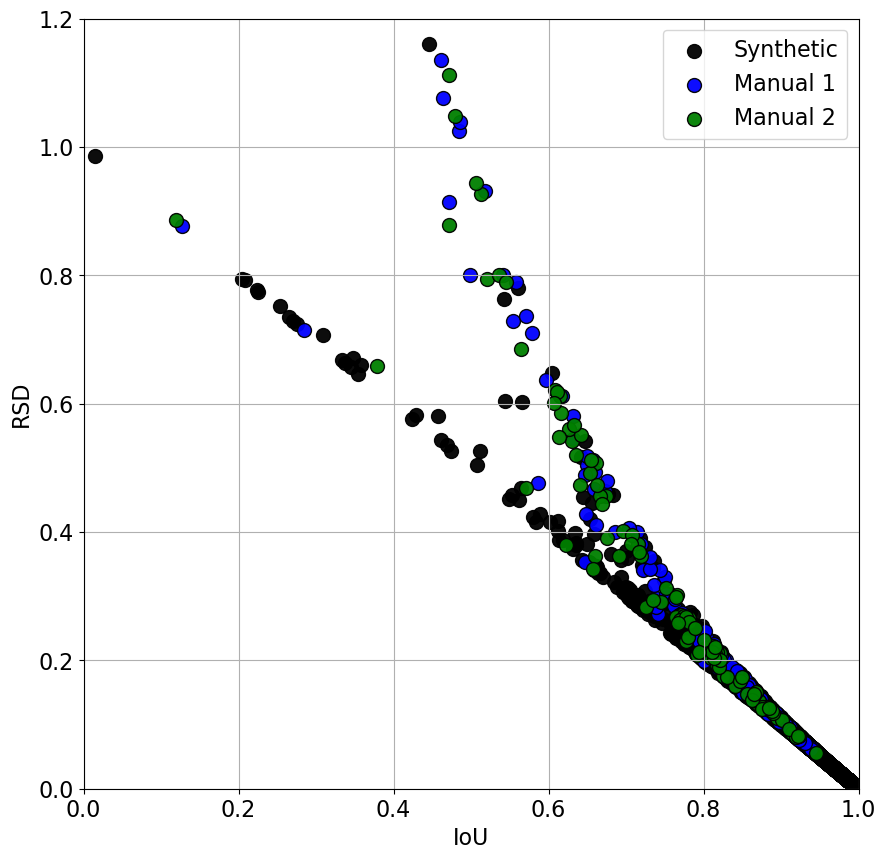}
  \caption{$RSD$ vs $IoU$, refer to Eq.~\ref{eq:iou}, computed for {$1.6\times10^4$} synthetic validation images (\textit{black points}) and 115 CME observations manually segmented by two independent operators (\textit{blue and green points}).}
  \label{fig:test_iou_nsd}
\end{figure*}

\begin{figure*}
  \centering
  \includegraphics[width=0.32\textwidth]{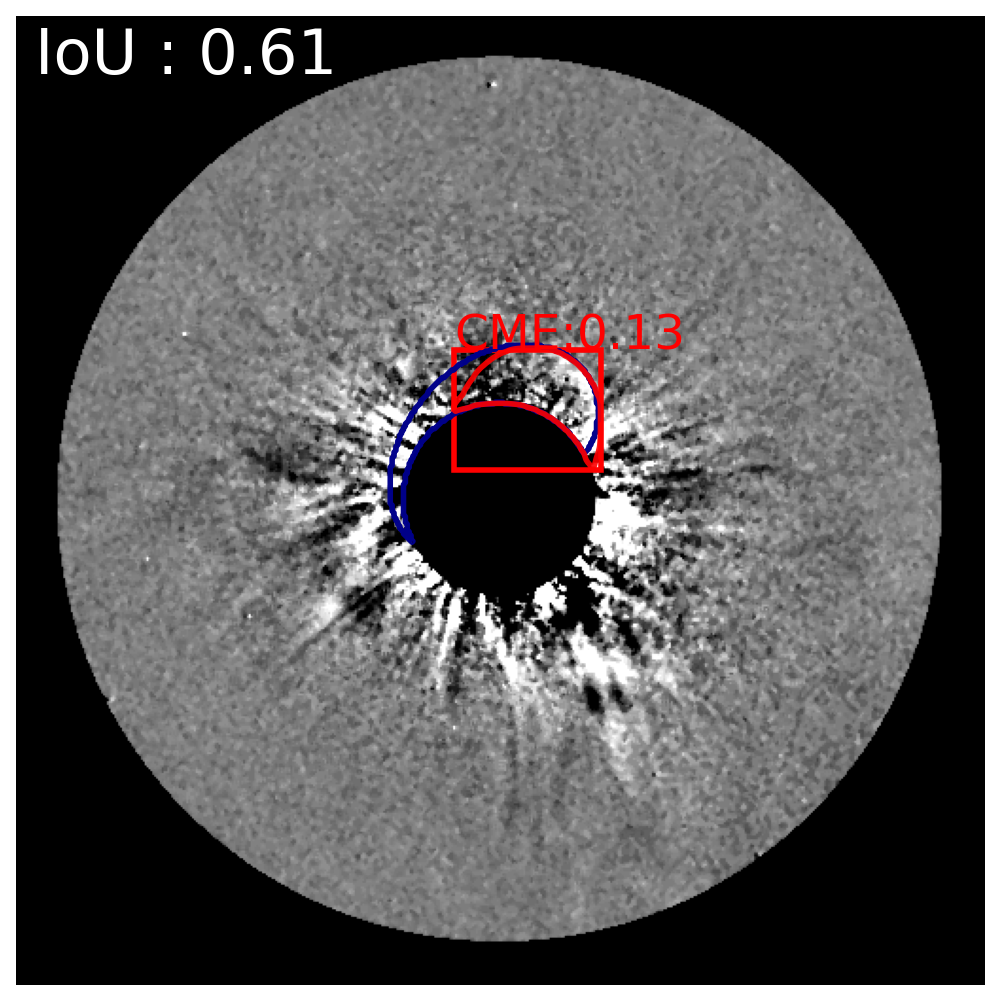}
  \includegraphics[width=0.32\textwidth]{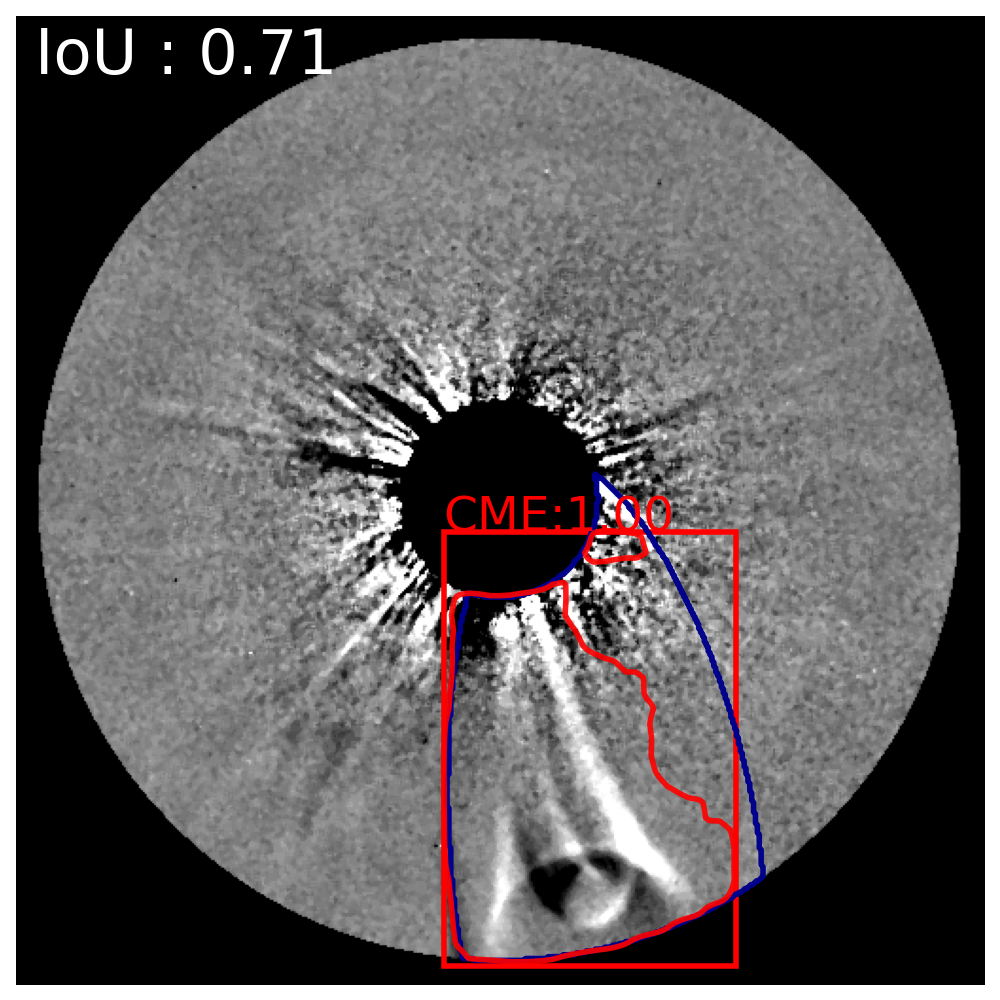}
  \includegraphics[width=0.32\textwidth]{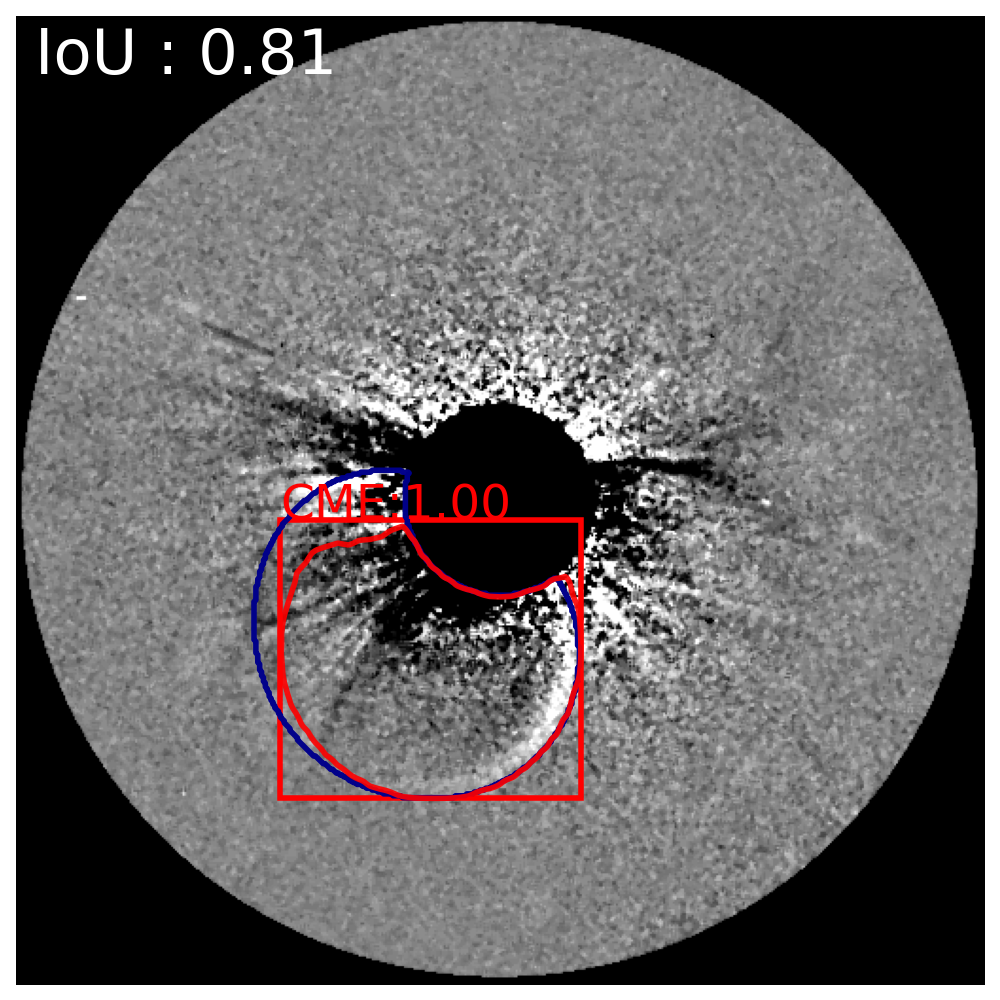}
  \includegraphics[width=0.32\textwidth]{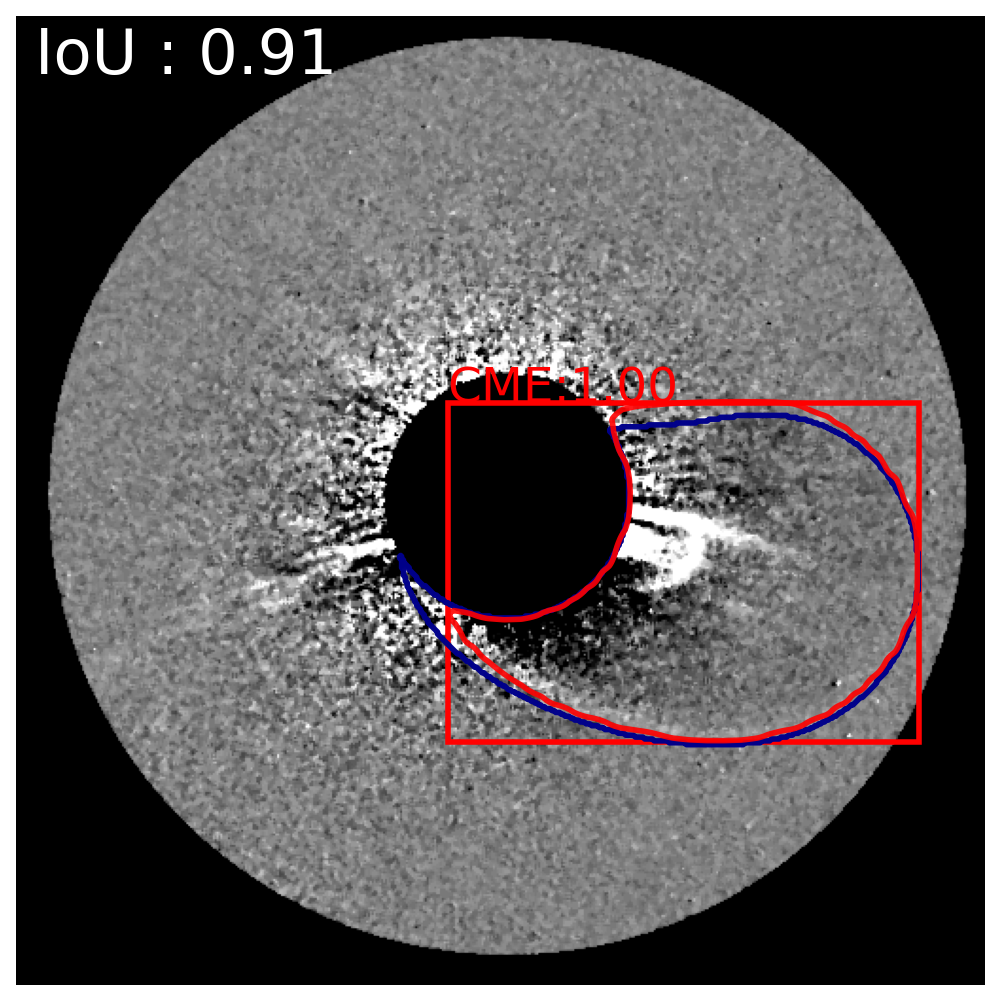}  
  \includegraphics[width=0.32\textwidth]{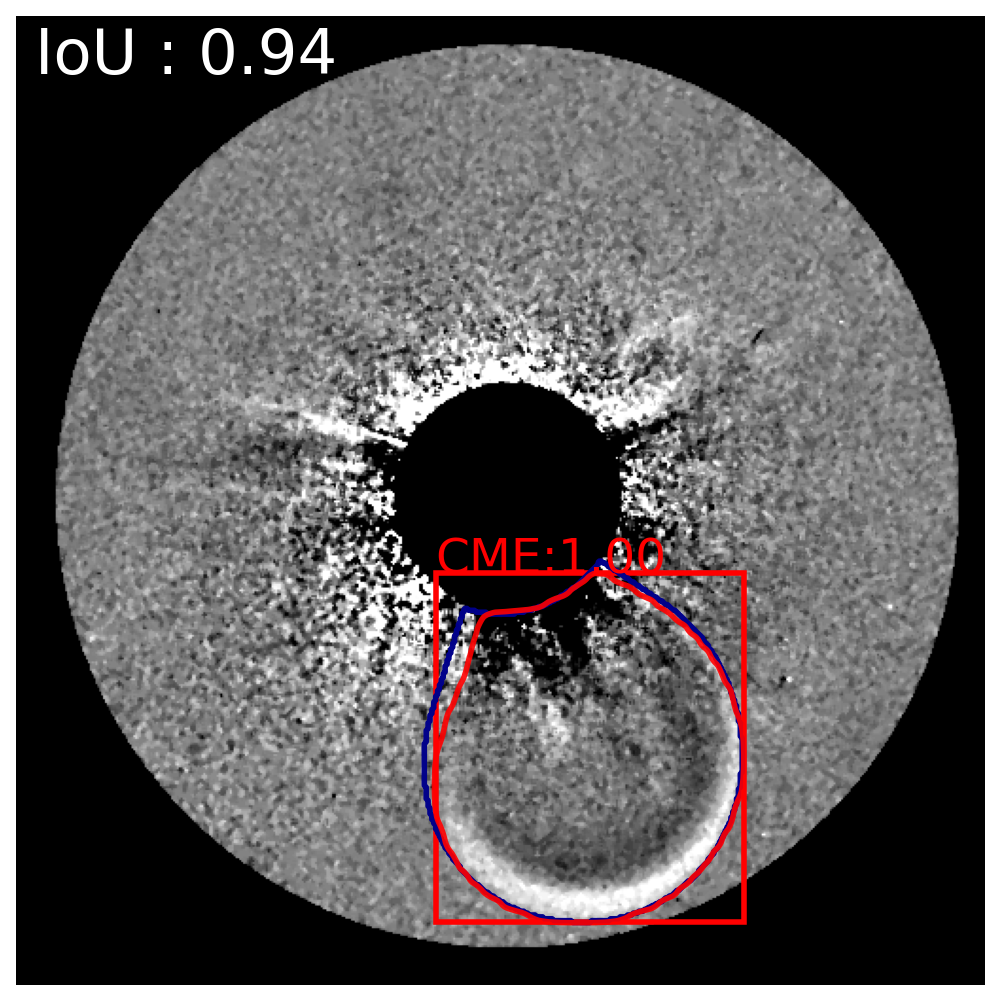}
  \includegraphics[width=0.32\textwidth]{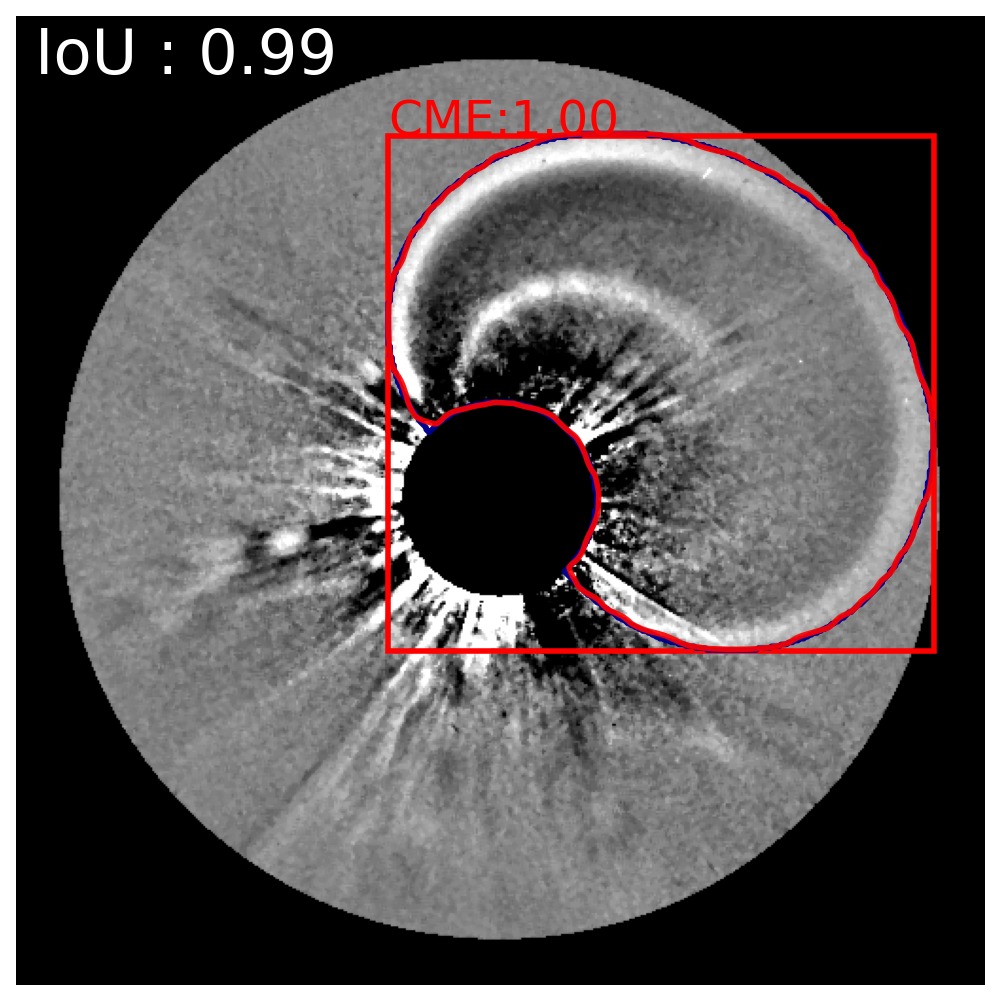}
  \caption{CME outer shell segmentation in synthetic validation images (\textit{red contour}) in comparison with the known true mask (\textit{blue contour}) for six values of $IoU$, ranging from poor (\textit{top left panel}) to excellent (\textit{bottom right panel}). Note that we only show the outlines of the masks and they may partially overlap.}
  \label{fig:res_syn_examples}
\end{figure*}

Fig.~\ref{fig:iou_vs_epoch} presents $IoU$ computed for $10^3$ random images of the synthetic validation dataset, at different stages during training. This is computed using the optimal mask threshold of Sec~\ref{sec:mask_gen} and, for each image, using only the output mask that best matches ${M}_{e0}$. Due to the latter, this $IoU$ does not measure false positive masks, while it is sensitive to false positive pixels within the selected mask. 
It can be seen that the initial $IoU$ is $\approx0.6$, which corresponds to the ResNet backbone pretrained on the COCO dataset (with no CME images). We also show in Fig.~\ref{fig:test_iou_nsd} the relationship between $IoU$ and the more intuitive $RSD$. Note that $RSD$ simply measures the non-overlapping area relative to the true mask area, thus depends on the true mask size. \textbf{For good performance, namely $IoU>0.8$ and $RSD<20\%$, the relationship is practically linear ($IoU=1-RSD$). For poor performance, there can be cases with $RSD>1$ if the total non-overleaping area is larger than the true mask area, see Eq.~\ref{eq:iou} and more details below in Sect. \ref{sec:result_real_obs}.} The initial and final validation $IoU$ found during training, 0.6 and 0.98, correspond to $RSD\approx40-65\%$ and $\approx2\%$, respectively. The median obtained with the trained Mask R-CNN on $1.6\times10^4$ validation images is 0.98, with $88\%$ ($1.4\times10^4$) of the masks having $IoU\ge0.95$ or $RSD\le5\%$. In Fig.~\ref{fig:res_syn_examples}, we present examples of detected masks along with the corresponding true mask for six values of $IoU$, ranging from poor to excellent.  Note that we only show the outline of the masks because they do not present holes or isolated patches.

\begin{figure*}
  \centering
  \includegraphics[width=0.49\textwidth]{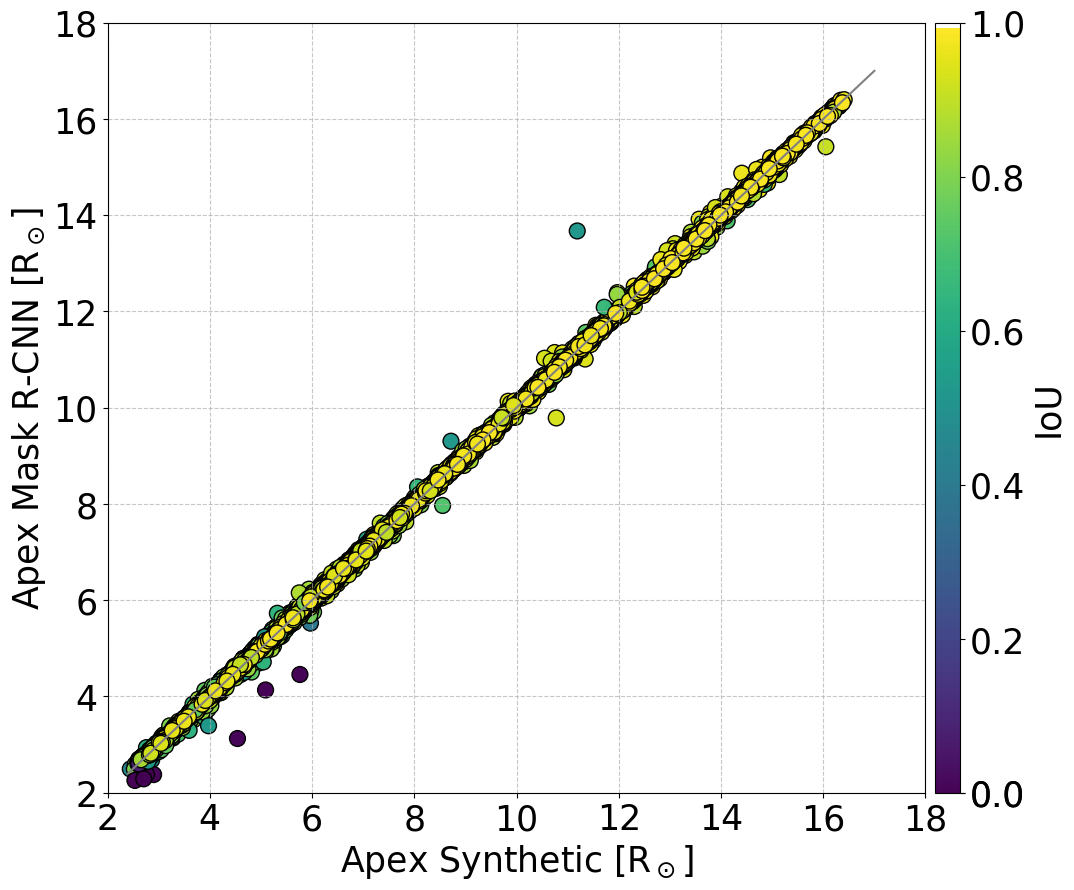}
  \includegraphics[width=0.49\textwidth]{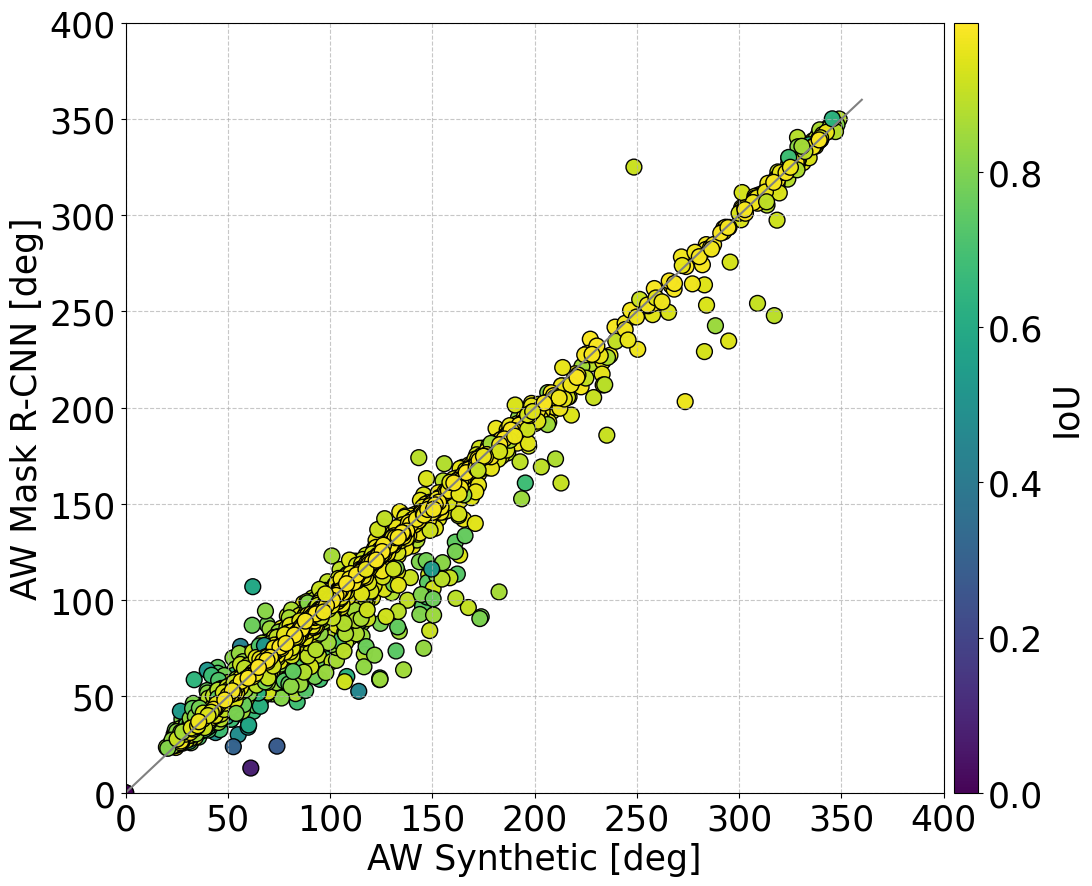}
  \includegraphics[width=0.49\textwidth]{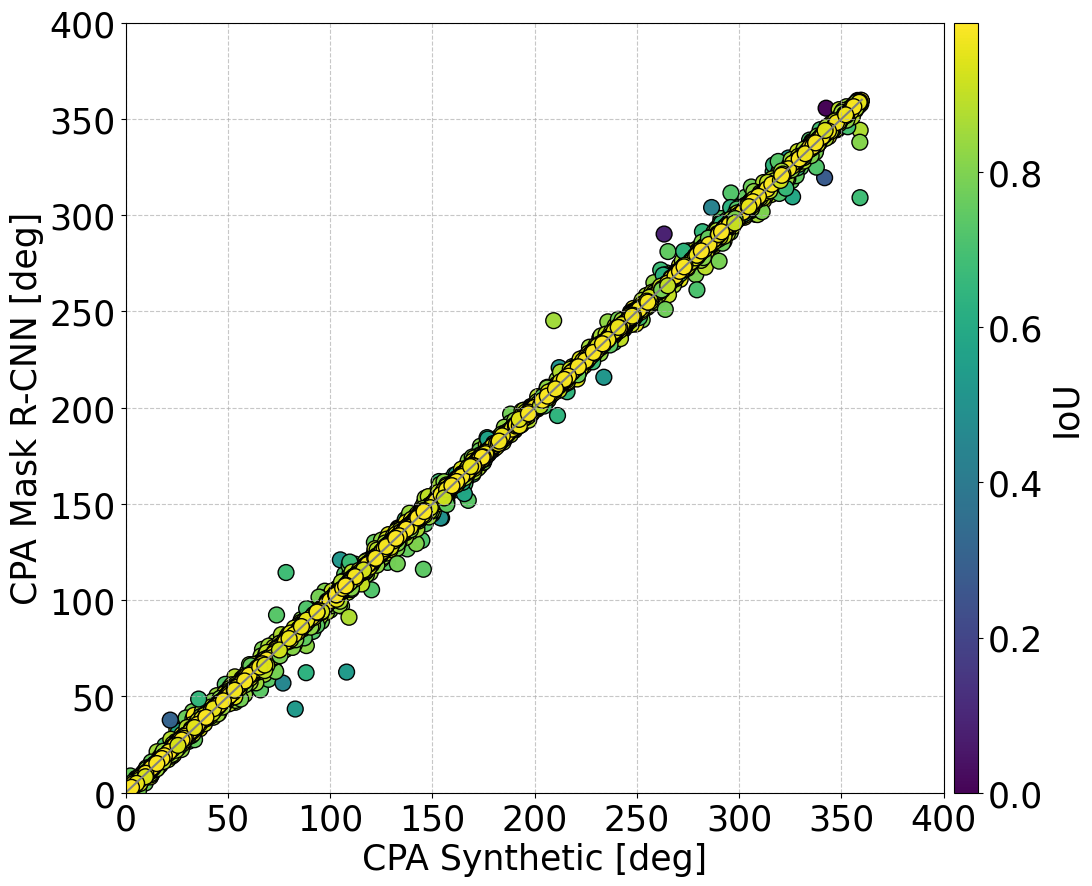}
  \caption{\textbf{Apex (\textit{top left}), AW (\textit{top right}) and CPA (\textit{bottom}) of the true masks (\textit{horizontal axes}) for {$1.6\times10^4$} synthetic validation images, versus the values for the masks obtained with our trained Mask R-CNN (\textit{vertical axes}). The color bar indicates the $IoU$ score of each image. The solid lines have slope one.}}
  \label{fig:res_validation_scatter}
\end{figure*}

\begin{figure*}
  \centering
  \includegraphics[width=1\textwidth]{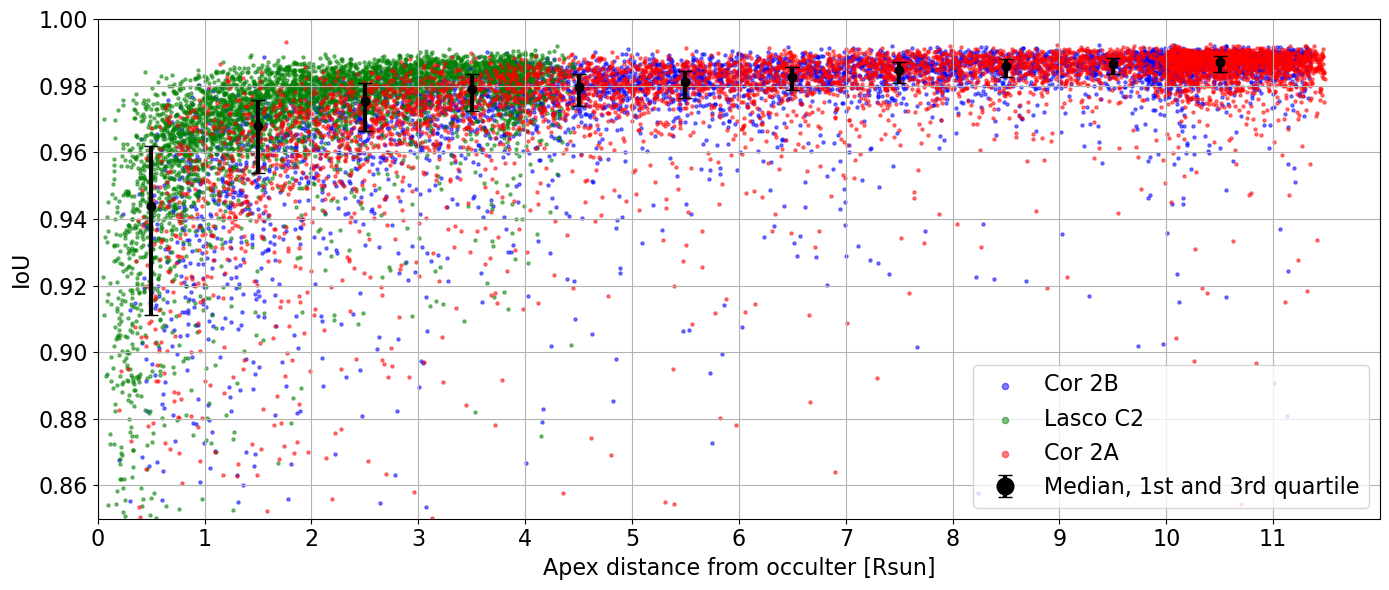}
  \caption{$IoU$ versus CME apex measured from the occulter for $1.6\times10^4$ validation synthetic images. LASCO C2 and COR2-A/B are indicated as green, red, and blue colors, respectively. Black dots and error bars indicate the median and 25th-75th percentiles for all cases in centered bins of size $\pm 0.5$~R$_{\odot}$.}
  \label{fig:res_syn_iou_vs_apex}
\end{figure*}

The variety of GCS shapes included in the synthetic dataset allows to study the performance of the detection with respect to various CME properties. \textbf{Fig.~\ref{fig:res_validation_scatter} presents the apex, AW and CPA of the true and inferred masks for the {$1.6\times10^4$} synthetic validation images. It can be seen that the agreement is better for apex and CPA than for AW, because the latter is very sensitive to the detailed mask shape near the occulter. The AW result suggest that the performance is worse for narrow than wide, halo-like CMEs. The median IoU for CMEs with AW$<30^\circ$ (122 cases) and $>300^\circ$ (132 cases) are 0.90 and 0.94, respectively.} \textbf{To study the variation of the inferred mask accuracy with respect to the CME height, we present in Fig.~\ref{fig:res_syn_iou_vs_apex} the $IoU$ versus CME apex, as measured from the occulter}. It can be seen that for lower heights the median detection quality and the spread degrade, with the lowest quartile going from $IoU\approx0.98$ ($RSD\approx2\%$) to $IoU\approx0.91$ ($RSD\approx9\%$). We associate this reduction to the increased contrast of the background coronal structures. Note that there can still be good detections at lower heights if the CME is well defined, namely it has a large differential signal with respect to the background. More importantly, this dependence is similar in all three instruments, even when their occulters have different radii, namely 1.7 and 2.5~R$_{\odot}$ for C2 and COR2, respectively. This suggests the relevant factor is related to the coronograph stray light rejection performance, which dominates the contrast near the occulter, over the intrinsic coronal structures.

\subsection{Performance on CME observations}
\label{sec:result_real_obs}

\begin{figure*}
  \centering
  \includegraphics[width=0.32\textwidth]{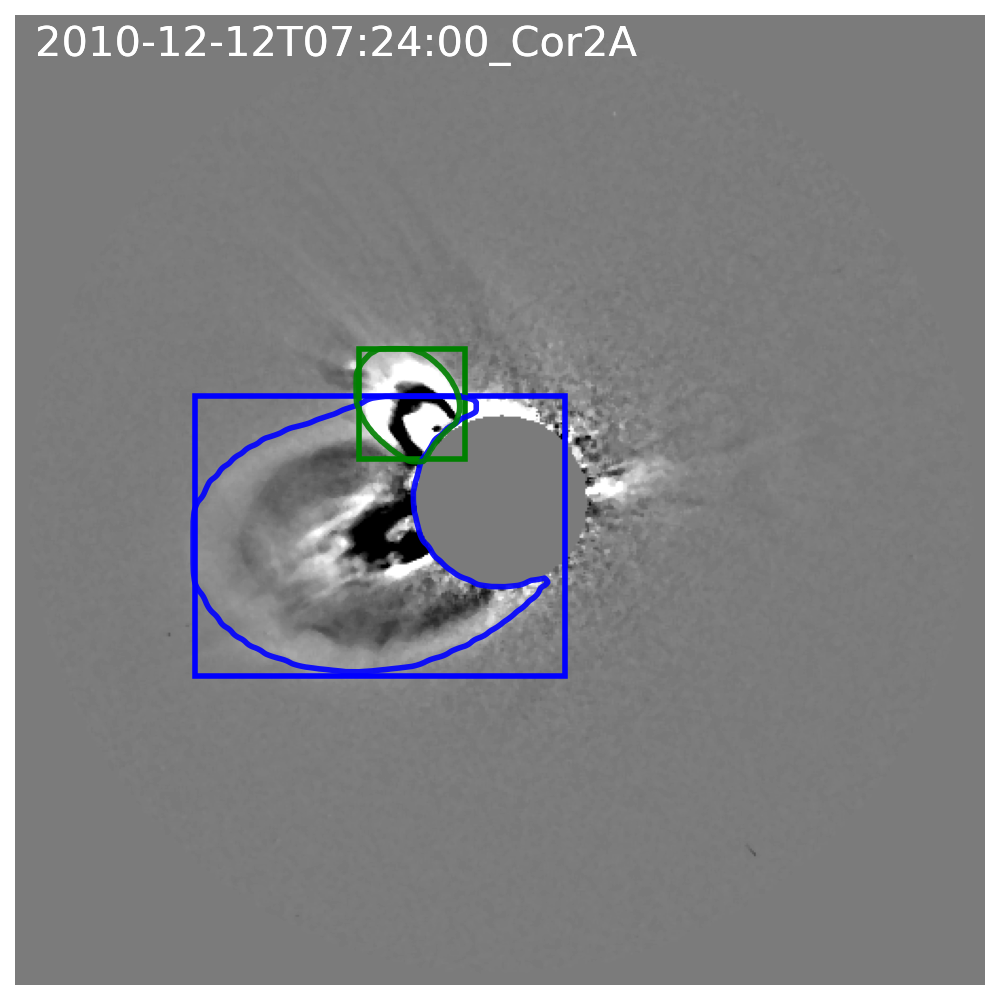}
  \includegraphics[width=0.32\textwidth]{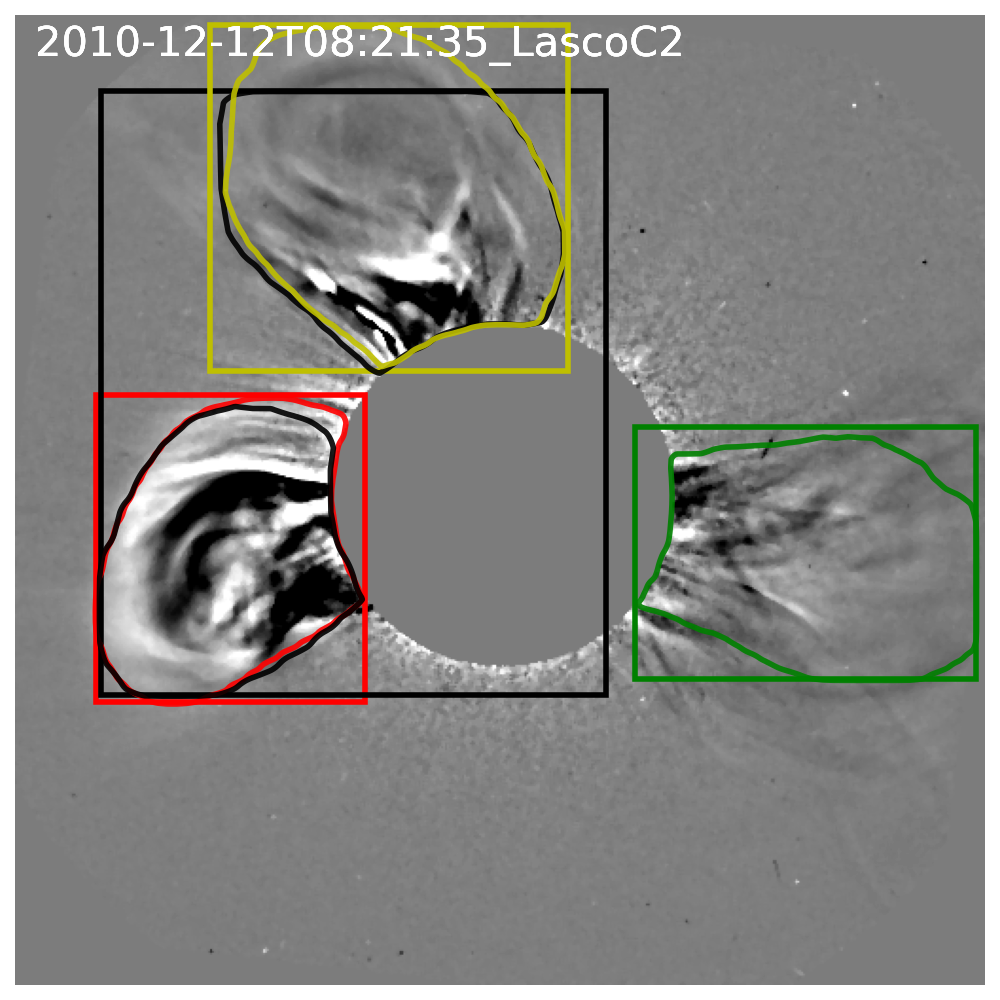}
  \includegraphics[width=0.32\textwidth]{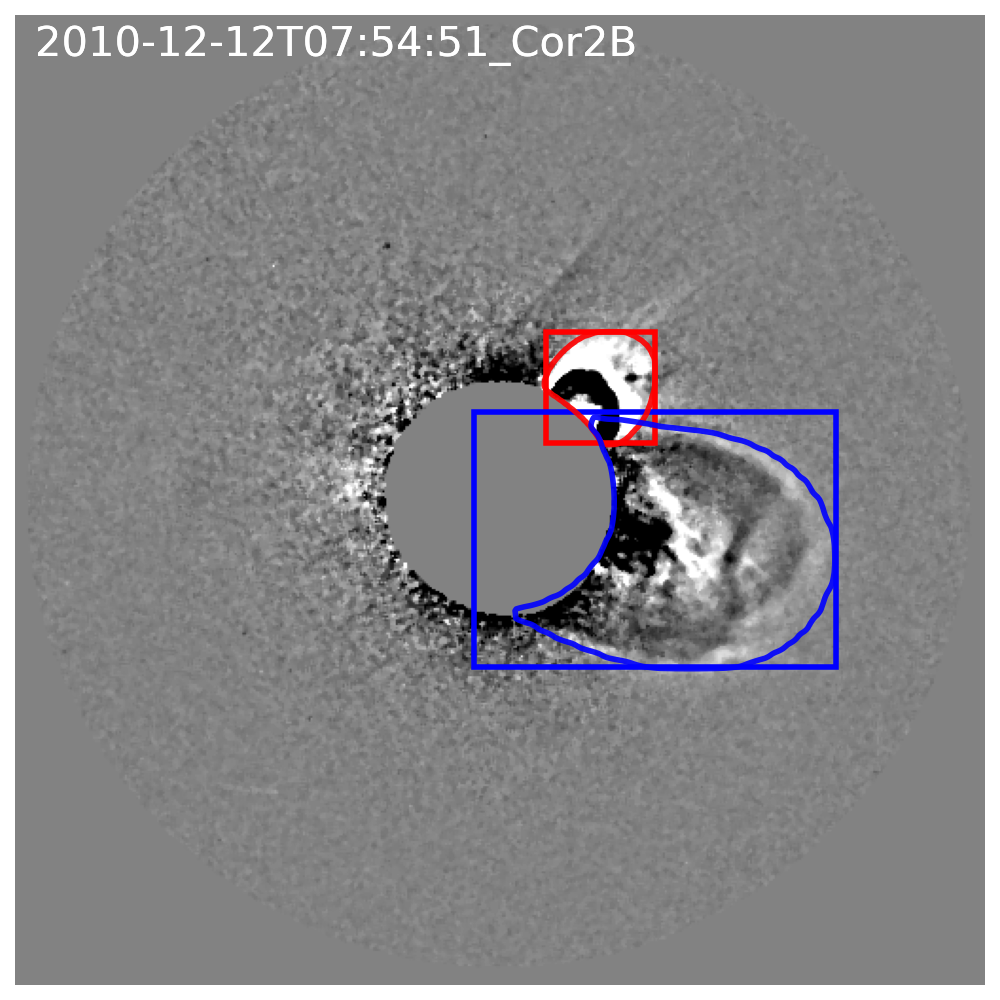}
  \caption{Segmentation of multiple CMEs observed by COR2-A (\textit{left}), LASCO C2 (\textit{middle}) and  COR2-B (\textit{right}). The outlines of all mask instances found  by our trained Mask R-CNN in each image independently are shown in different colors. Note that some outlines partially overlap.}
  \label{fig:obs_mult_cme}
\end{figure*}

\begin{figure*}
  \centering
  \includegraphics[width=0.98\textwidth]{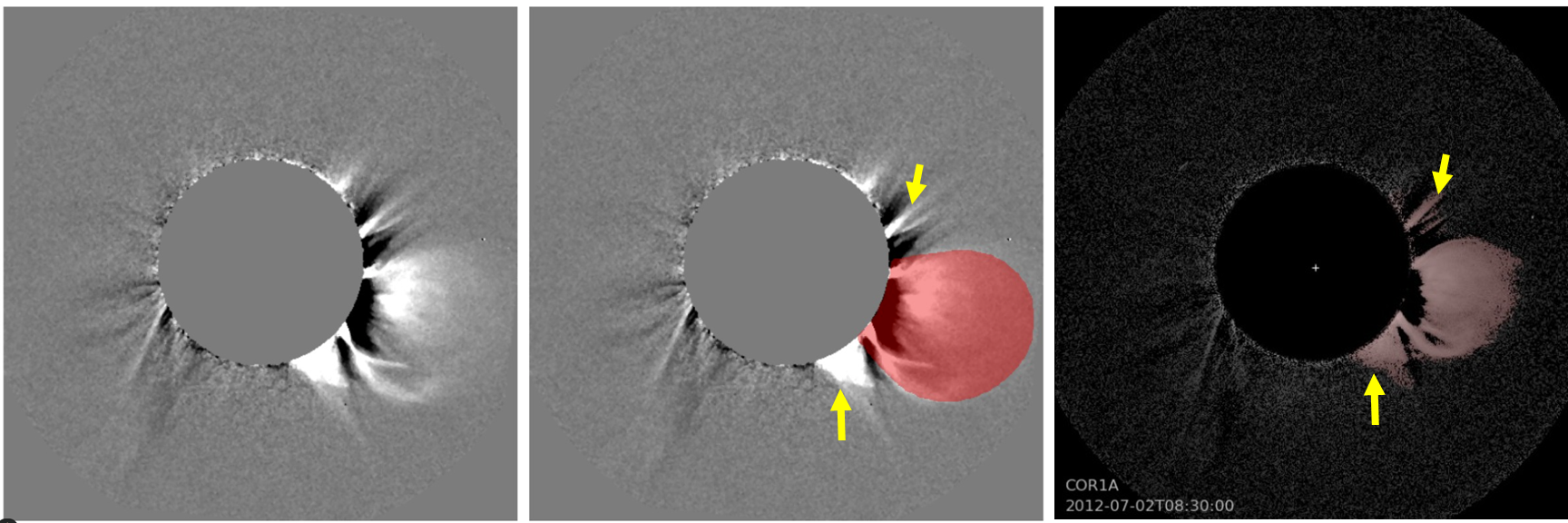}
  \caption{Segmentation comparison  for the CME observed by COR1-A on 2012/07/02 at 08:30:00 (\textit{left}). Note that our model-based approach \textit{(middle)} produces simpler shapes than \cite{shan2024} \textit{(right)}, but is able to better distinguish between the main CME bulk and nearby fast moving material (\textit{yellow arrows}).}
  \label{fig:ours_vs_camelII}
\end{figure*}

Mask R-CNN can identify multiple instances of the CME class. When applied to coronagraphic observations, this improves the detection of multiple CMEs in a single differential image, even when CMEs are close to each other (see Fig. \ref{fig:obs_mult_cme}), and reduces the false detection of fast moving background material. The latter is particularly challenging to achieve using traditional segmentation algorithms or other neural methods, which do not incorporate morphological constraints. For example, Fig.~\ref{fig:ours_vs_camelII} shows the segmentations obtained by our method and by \cite{shan2024} as presented in their public catalog\footnote{Available at \url{http://aso-s.pmo.ac.cn/feature/\#/CME\_2d\_catalog}}. Note that our model-based training approach produces simpler shapes, which may not be representative of complex CMEs, but can better distinguish between the main CME bulk and other fast moving material based only on their morphology, i.e., with no further kinematic information. 

\begin{figure*}
  \centering
  \includegraphics[width=0.24\textwidth, trim=2cm 0cm 13.5cm 1cm, clip]{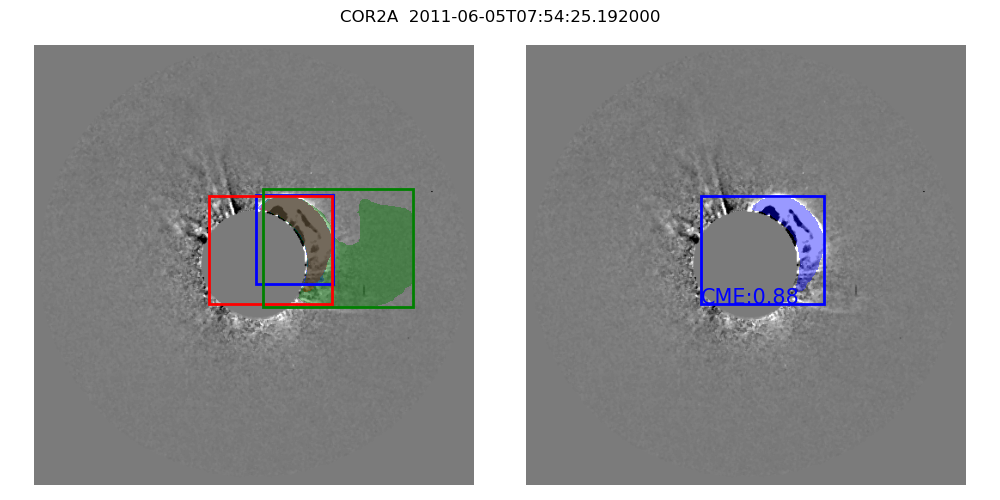}
  \includegraphics[width=0.24\textwidth, trim=2cm 0cm 13.5cm 1cm, clip]{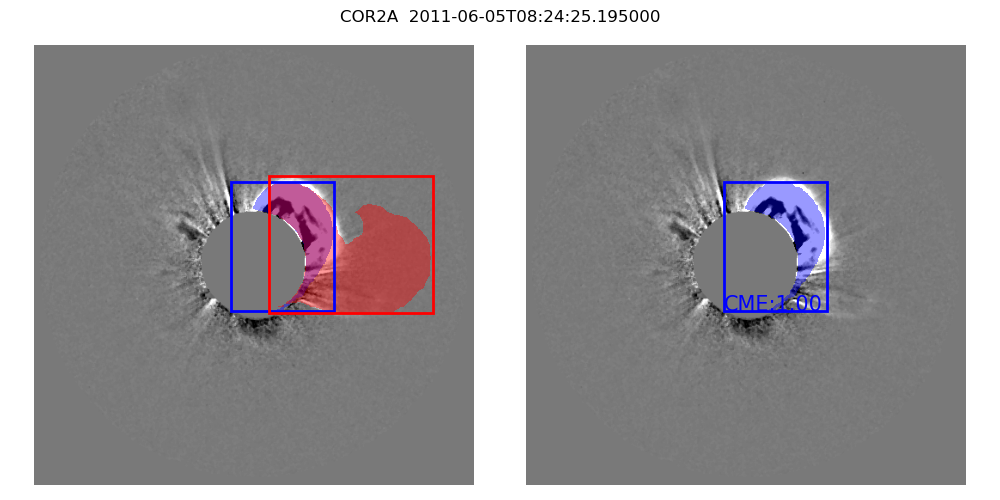}
  \includegraphics[width=0.24\textwidth, trim=2cm 0cm 13.5cm 1cm, clip]{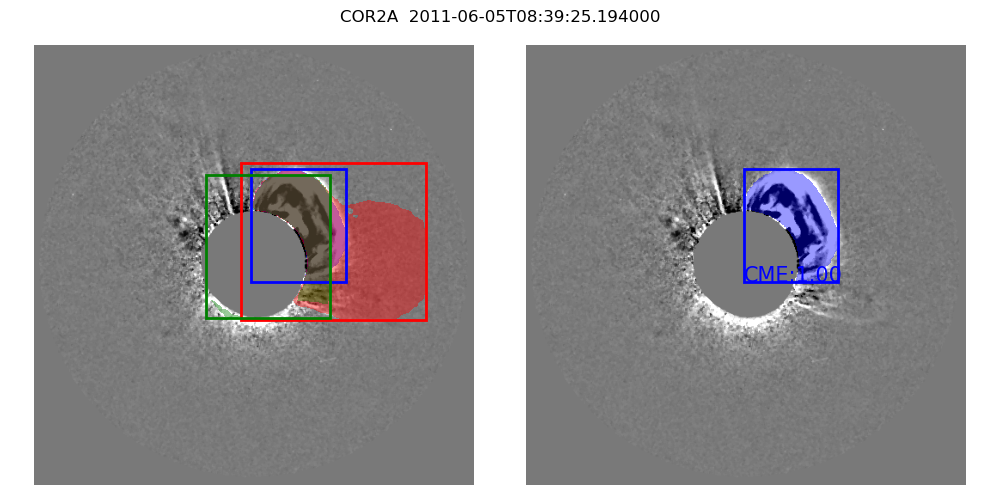}
  \includegraphics[width=0.24\textwidth, trim=2cm 0cm 13.5cm 1cm, clip]{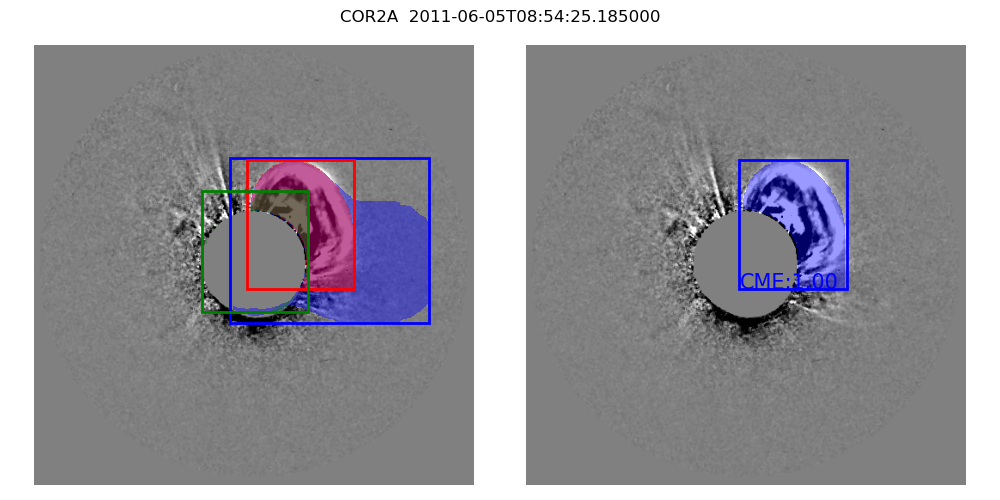}  
  \includegraphics[width=0.24\textwidth, trim=13.5cm 0cm 2cm 1cm, clip]{FMwLASCO201106056_seq_COR2A.png}
  \includegraphics[width=0.24\textwidth, trim=13.5cm 0cm 2cm 1cm, clip]{FMwLASCO201106057_seq_COR2A.png}
  \includegraphics[width=0.24\textwidth, trim=13.5cm 0cm 2cm 1cm, clip]{FMwLASCO201106058_seq_COR2A.png}
  \includegraphics[width=0.24\textwidth, trim=13.5cm 0cm 2cm 1cm, clip]{FMwLASCO201106059_seq_COR2A.png}   
  \caption{Mask instance selection by CME tracking. We show in different colors all the masks found by Mask R-CNN (\textit{panels in the top row}) for the CME observed by Cor2A on 2011/06/05 at 07:54:25, 08:24:25, 08:39:25 and 08:54:25 (\textit{from left to right}). The selected masks are given in blue in the bottom row panels.}
  \label{fig:res_cme_tracking}
\end{figure*}

If the time series of images for a CME event is available, the expected temporal variation of basic CME morphological properties can be used to select one of the multiple inferred masks in each image, minimizing false positive masks. The result of applying the procedure described in Sect.~\ref{sec:instance_selec} is shown in Fig.~\ref{fig:res_cme_tracking}.

\begin{figure*}
  \centering
  \includegraphics[width=0.85\textwidth]{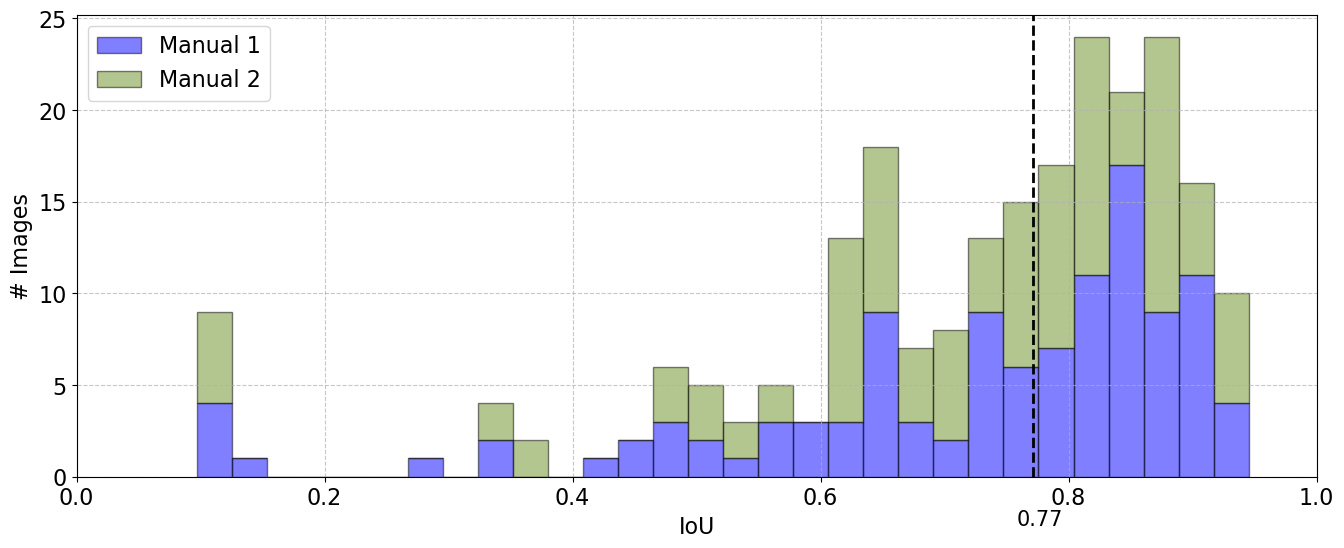}
  \caption{Histogram of IoU between masks inferred by our method and manual operators 1 (\textit{blue}) and 2 (\textit{green}), for 115 CME observations acquired by COR2-A, COR2-B and LASCO C2. The dashed vertical line indicates the overall median. See the text for extra details.}
  \label{fig:res_paper_events_iou_hist}
\end{figure*}

\begin{figure*}
  \centering
    \includegraphics[width=0.32\textwidth]{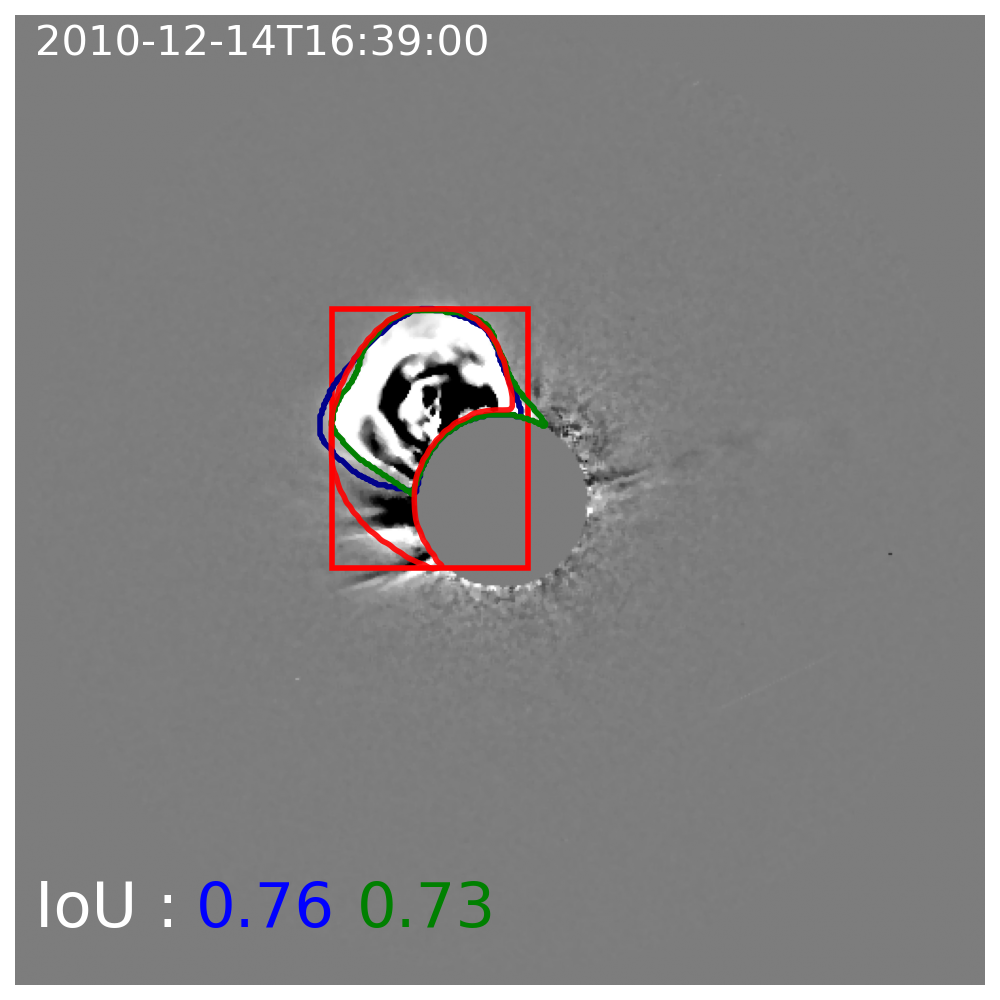} 
    \includegraphics[width=0.32\textwidth]{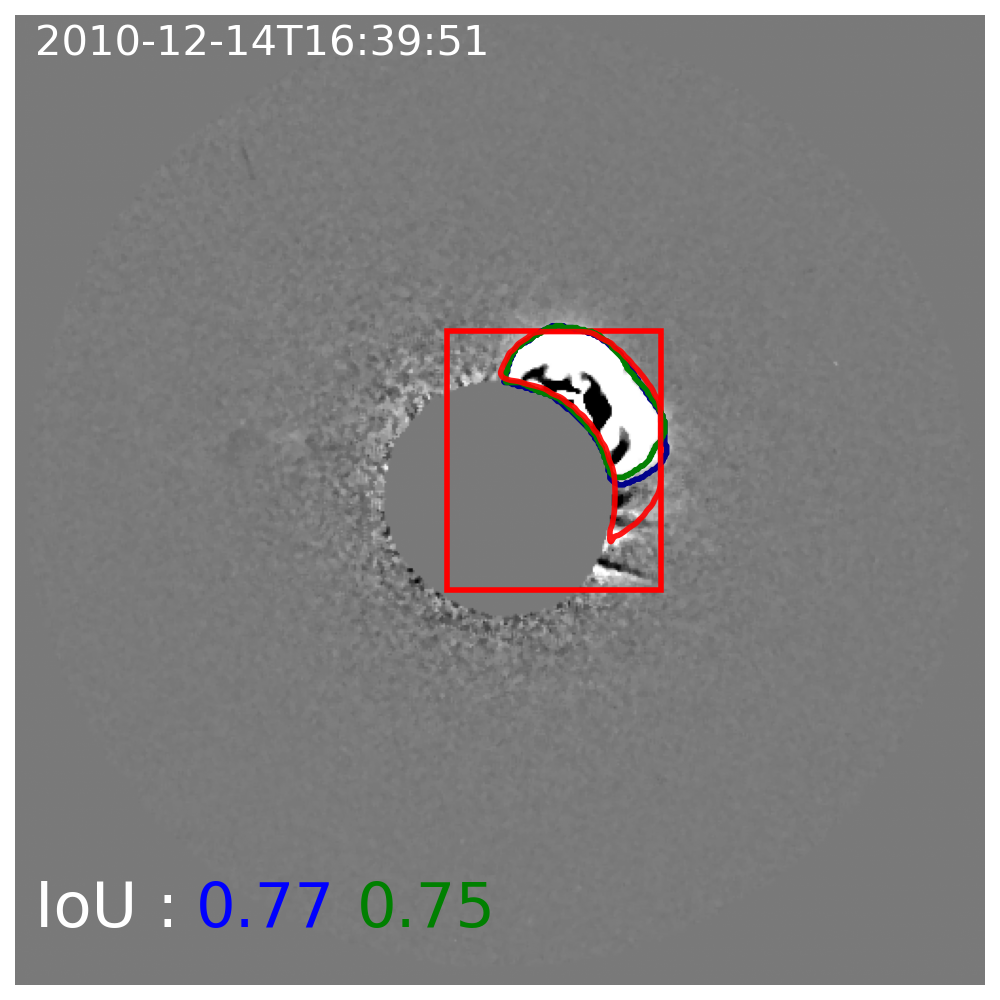} 
    \includegraphics[width=0.32\textwidth]{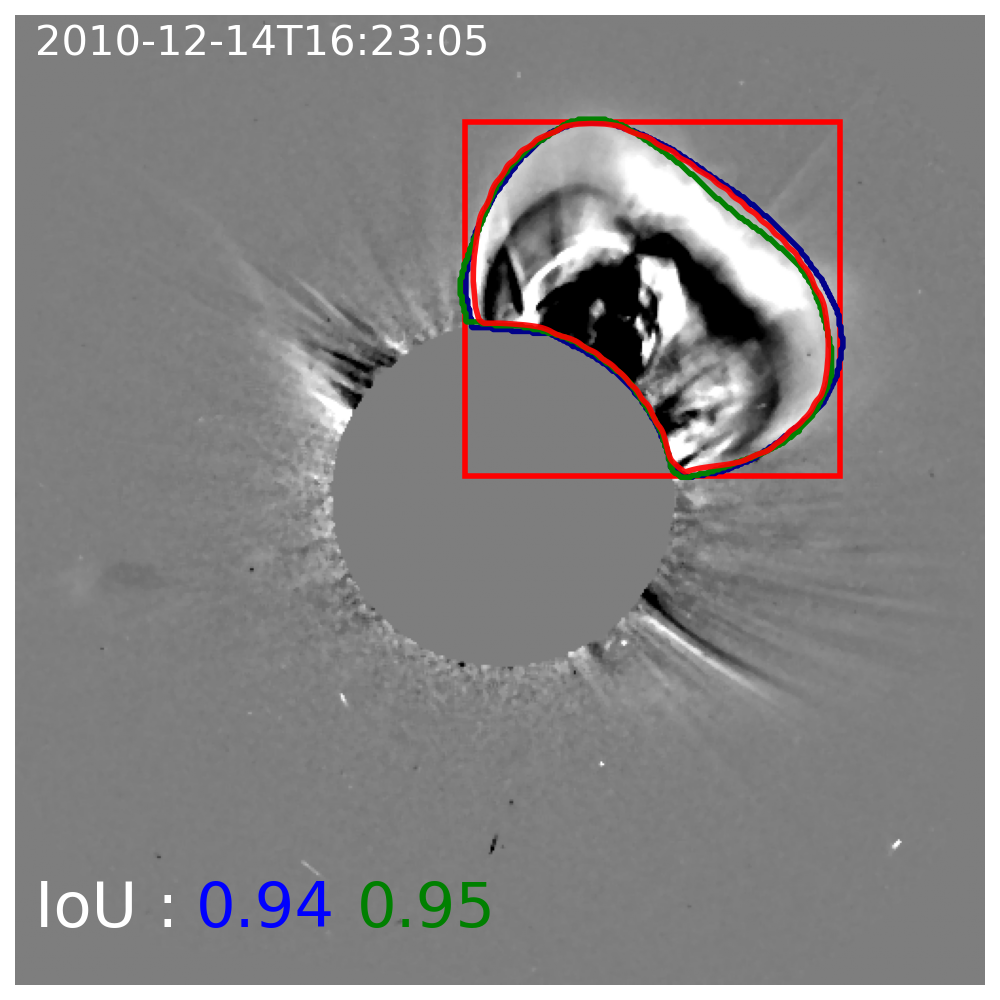}
    \includegraphics[width=0.32\textwidth]{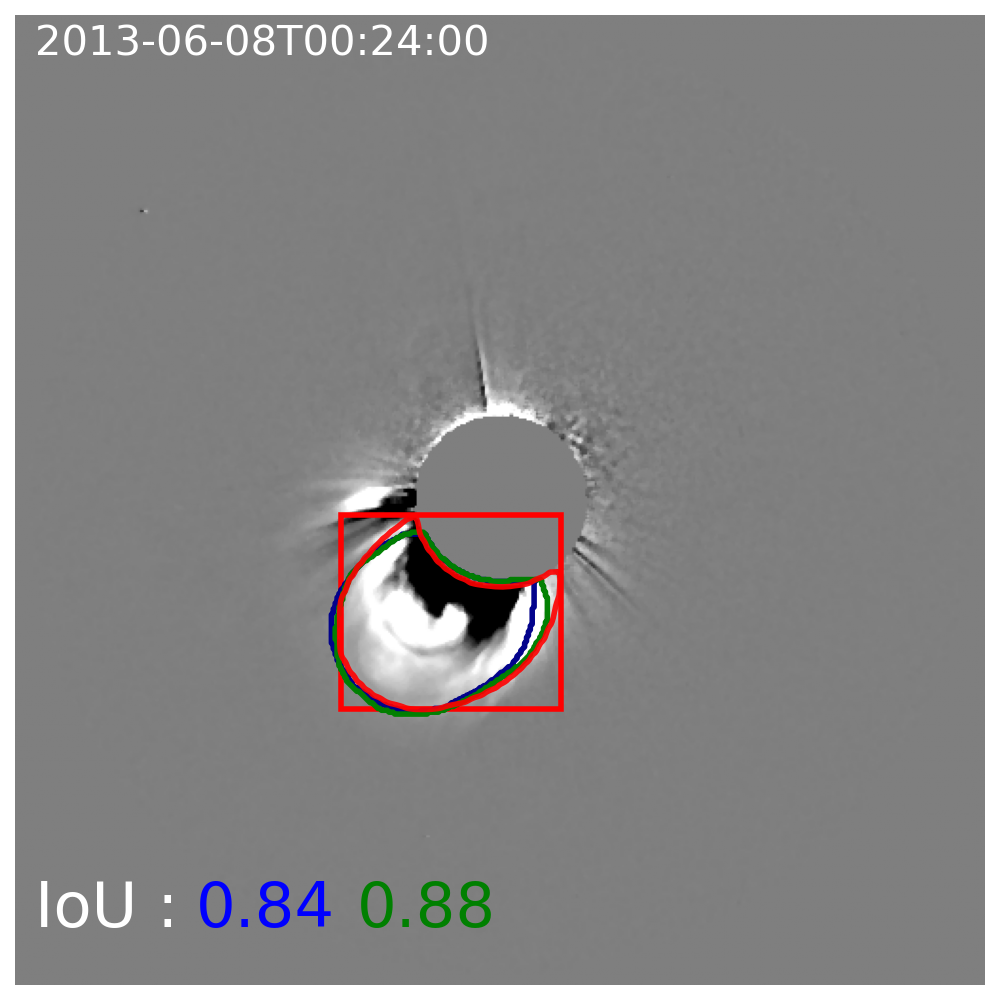}
    \includegraphics[width=0.32\textwidth]{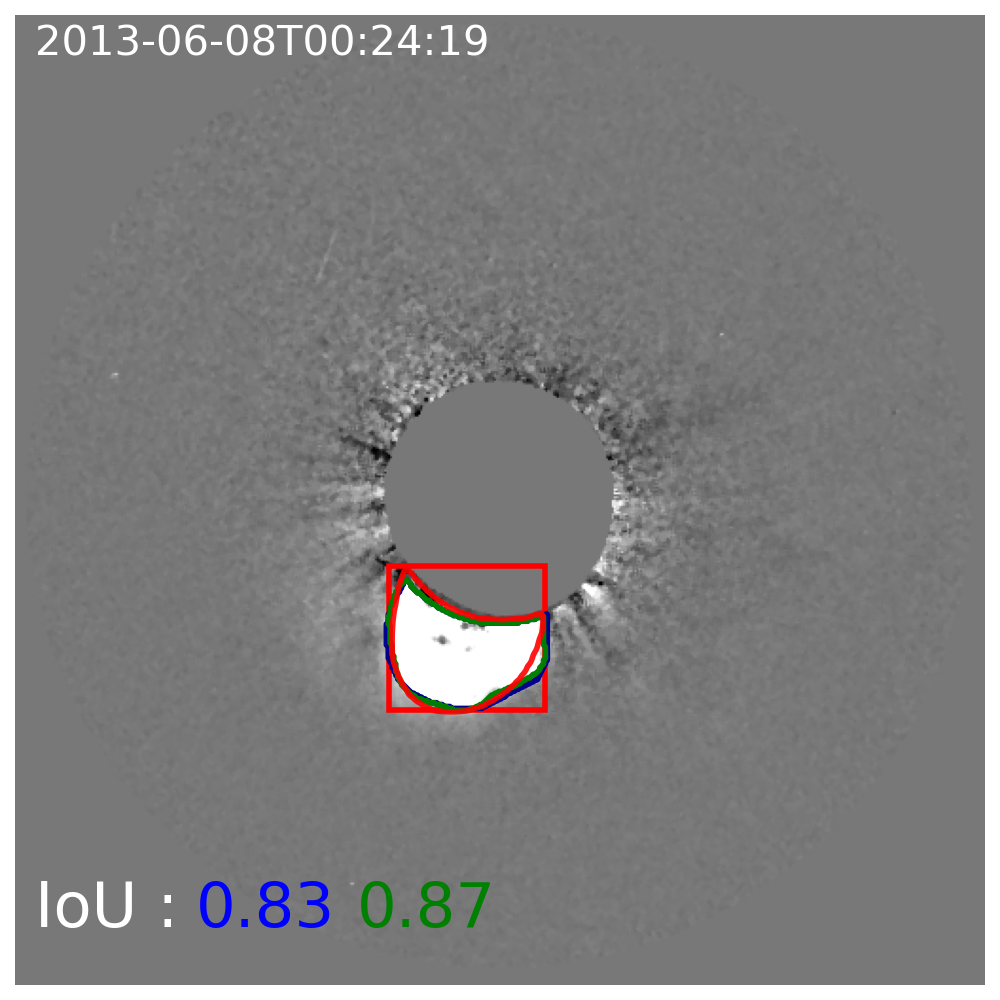}
    \includegraphics[width=0.32\textwidth]{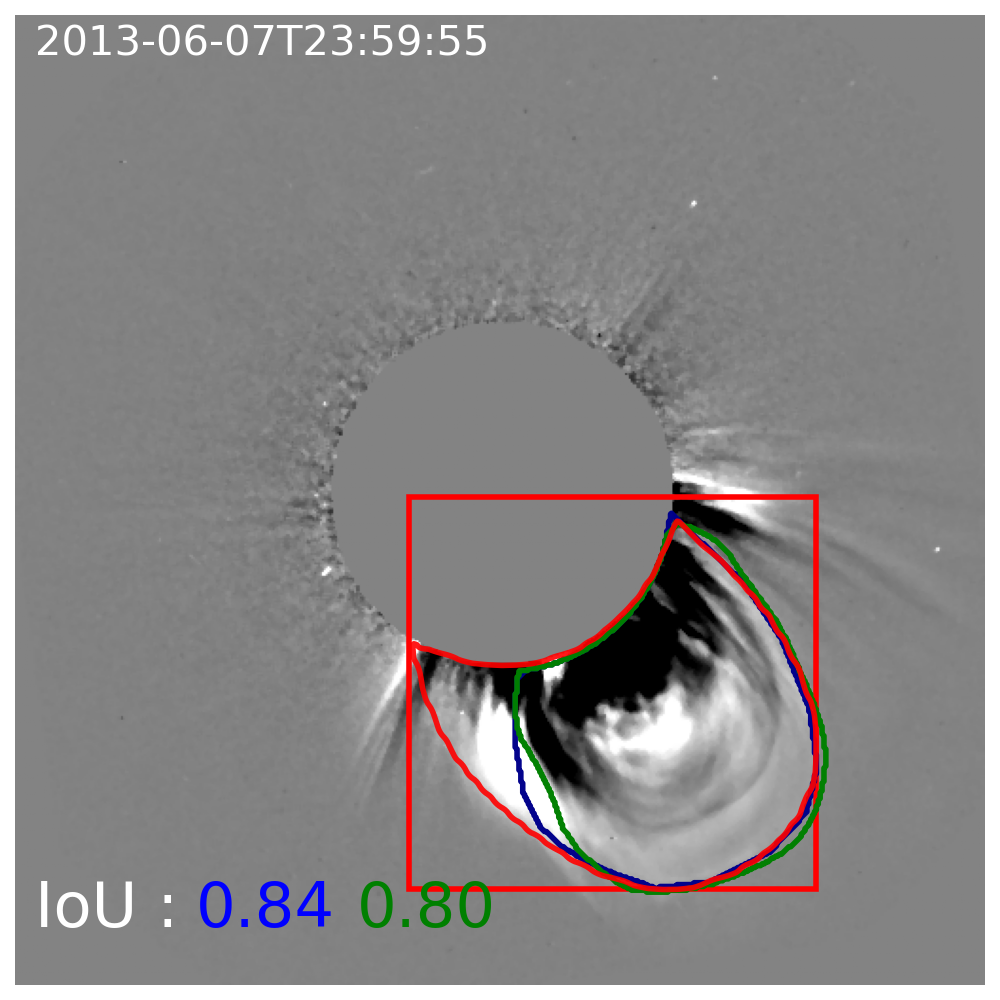}
    \includegraphics[width=0.32\textwidth]{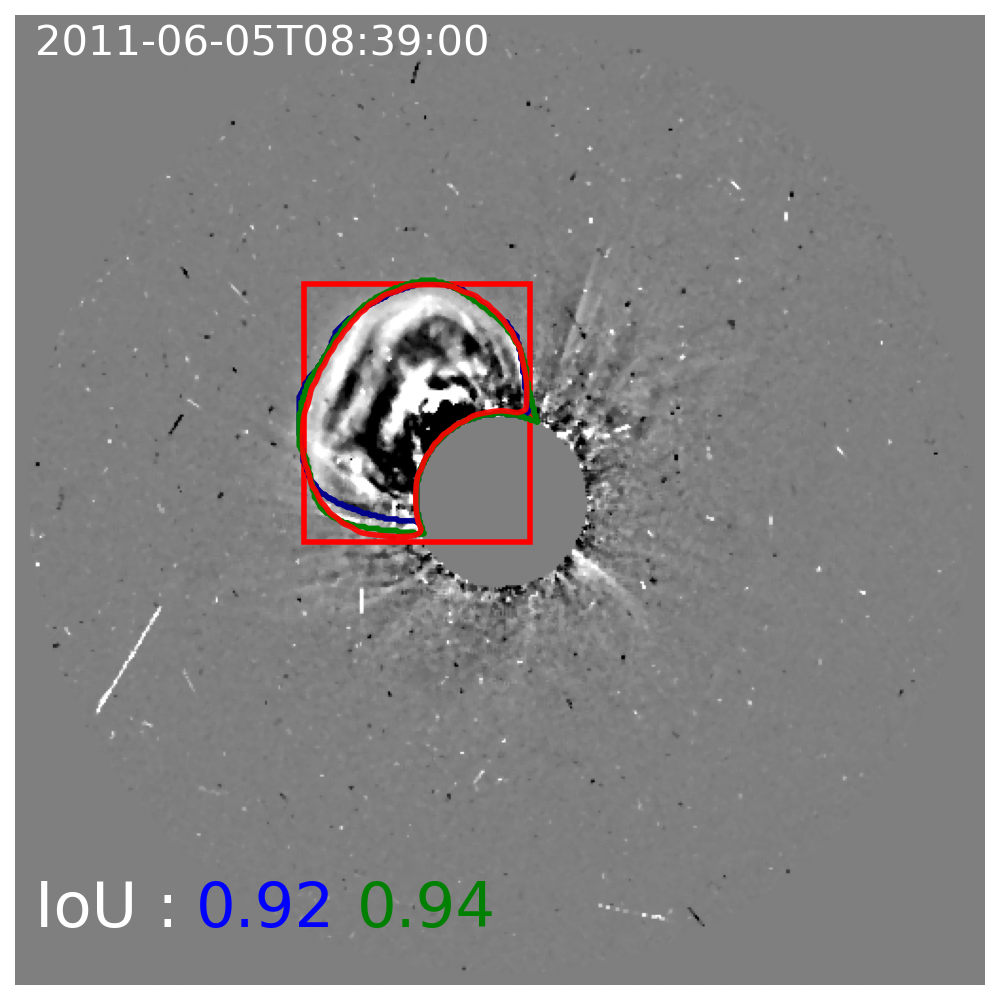}
    \includegraphics[width=0.32\textwidth]{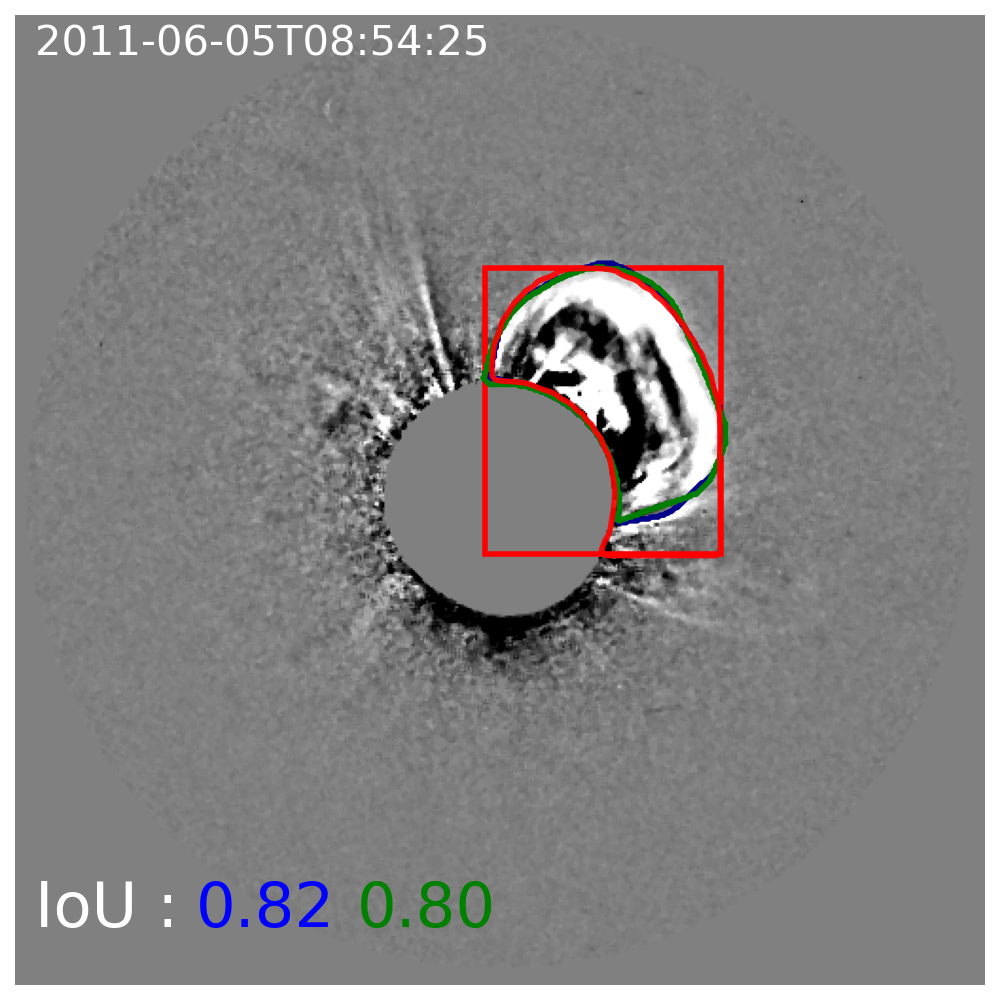}
    \includegraphics[width=0.32\textwidth]{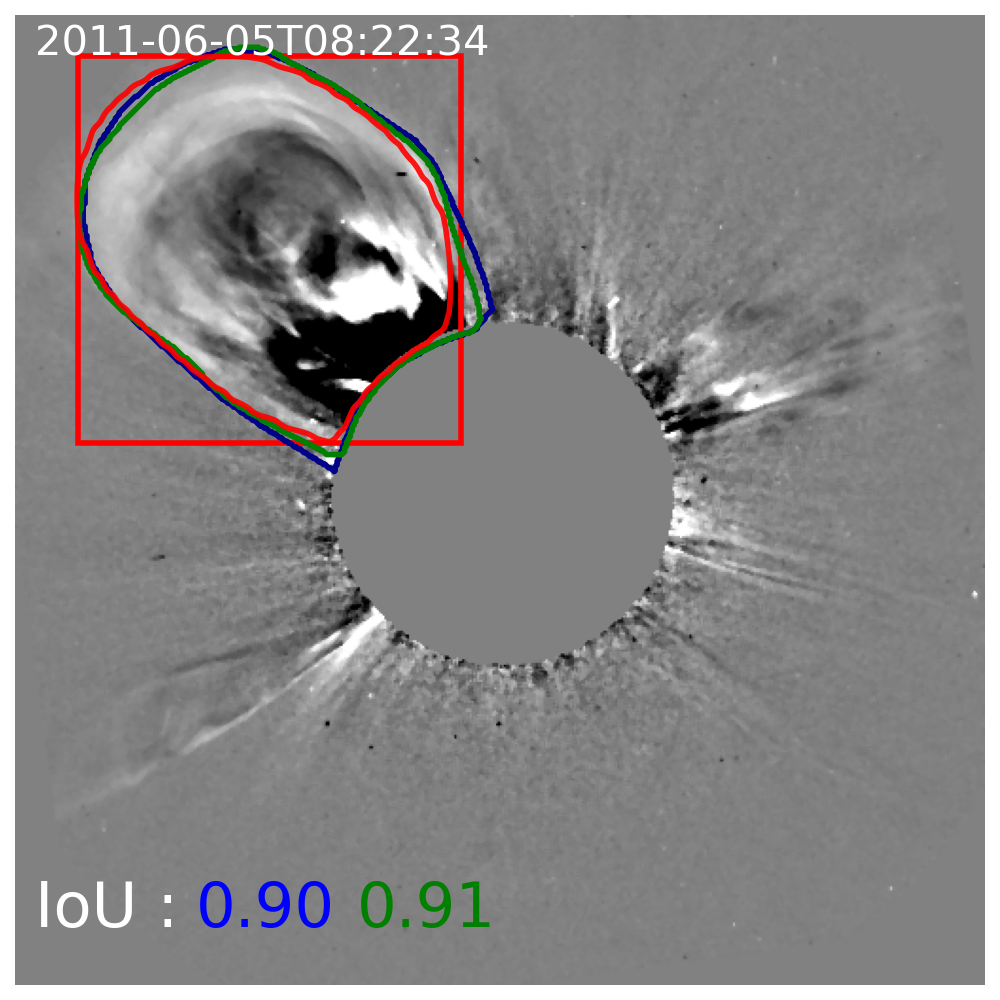}
    \includegraphics[width=0.32\textwidth]{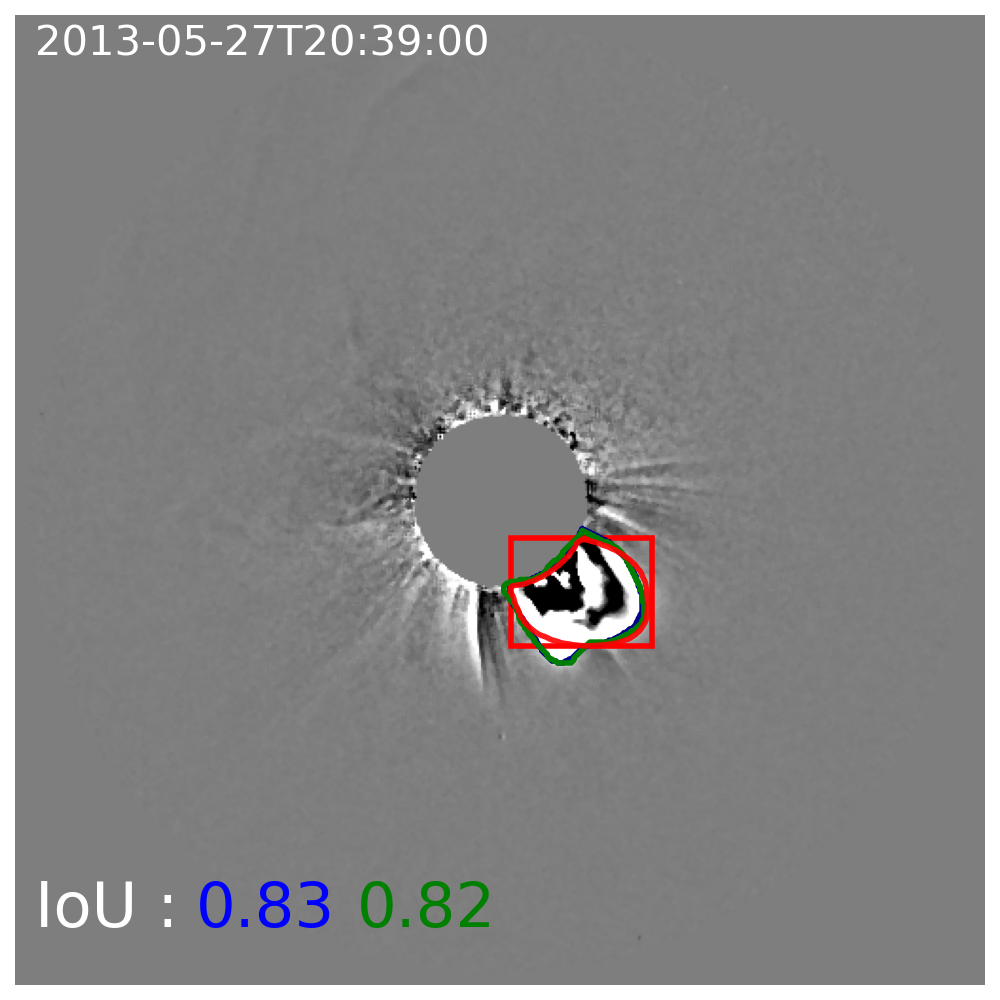} 
    \includegraphics[width=0.32\textwidth]{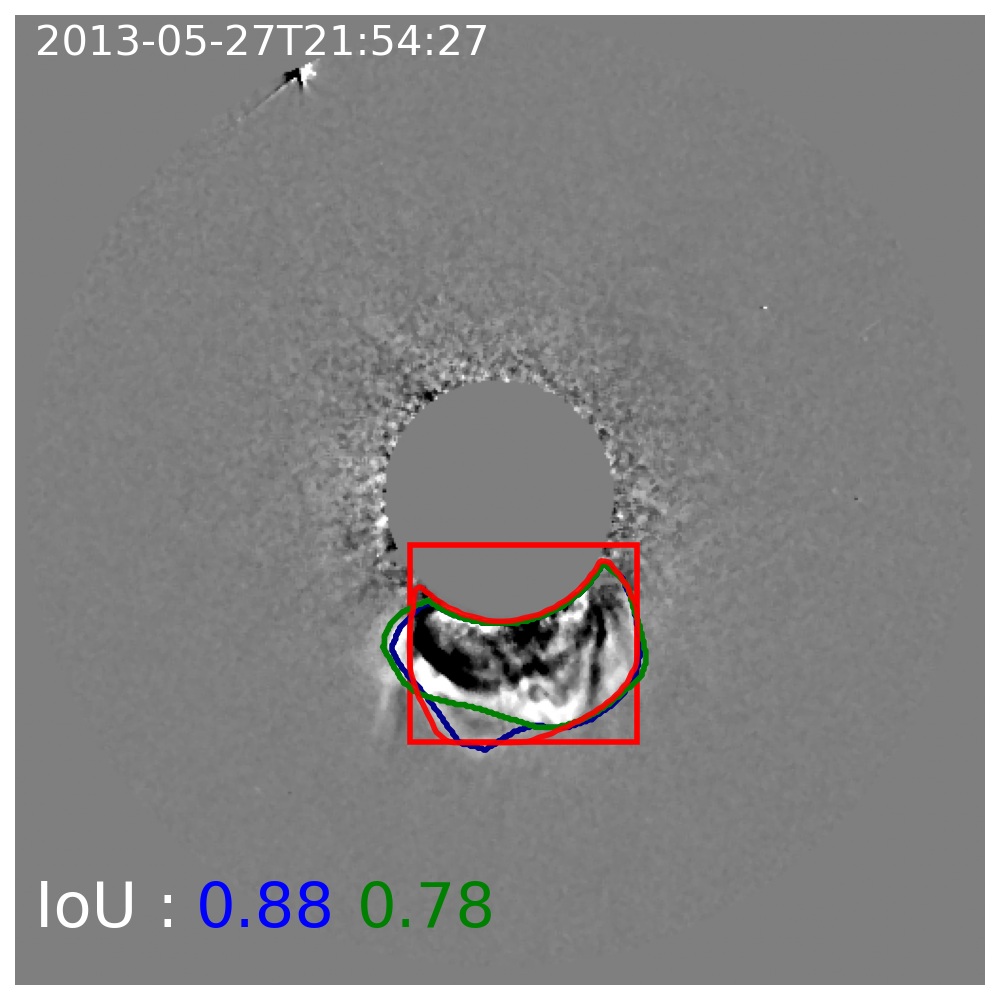} 
    \includegraphics[width=0.32\textwidth]{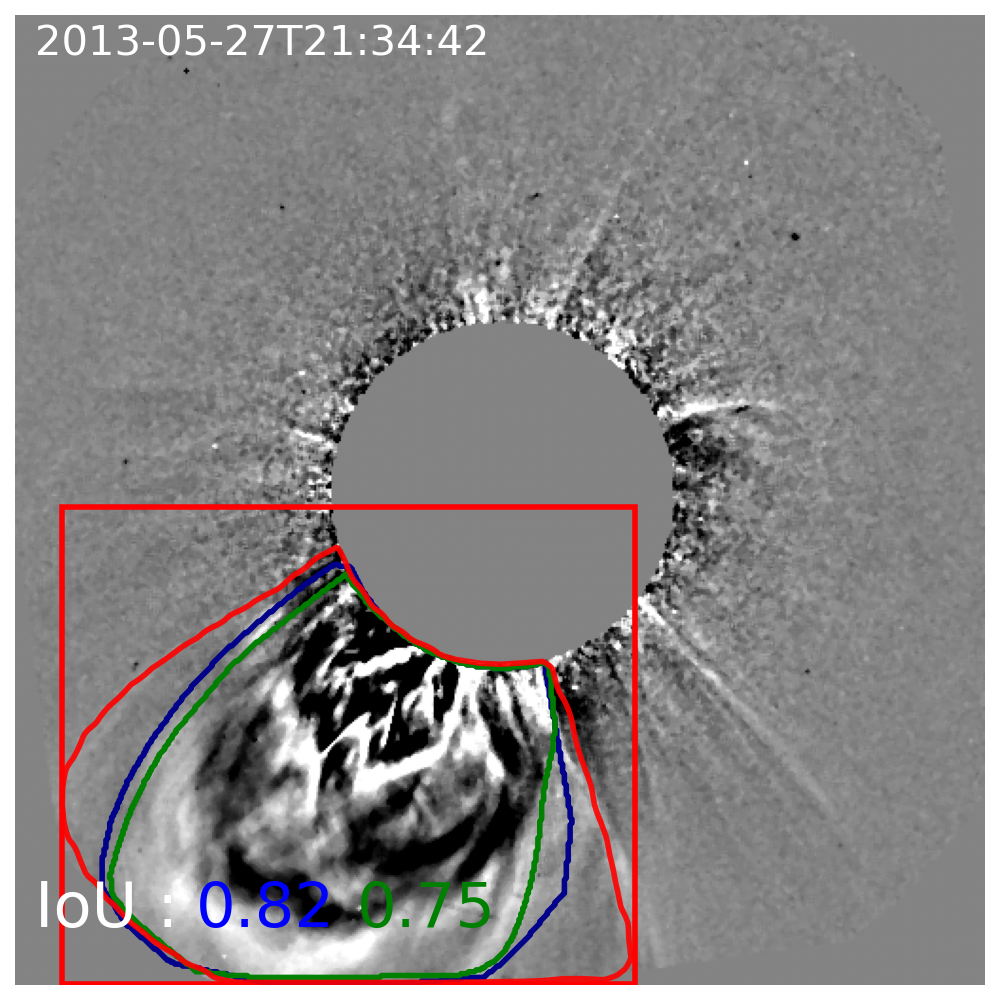}
    \includegraphics[width=0.32\textwidth]{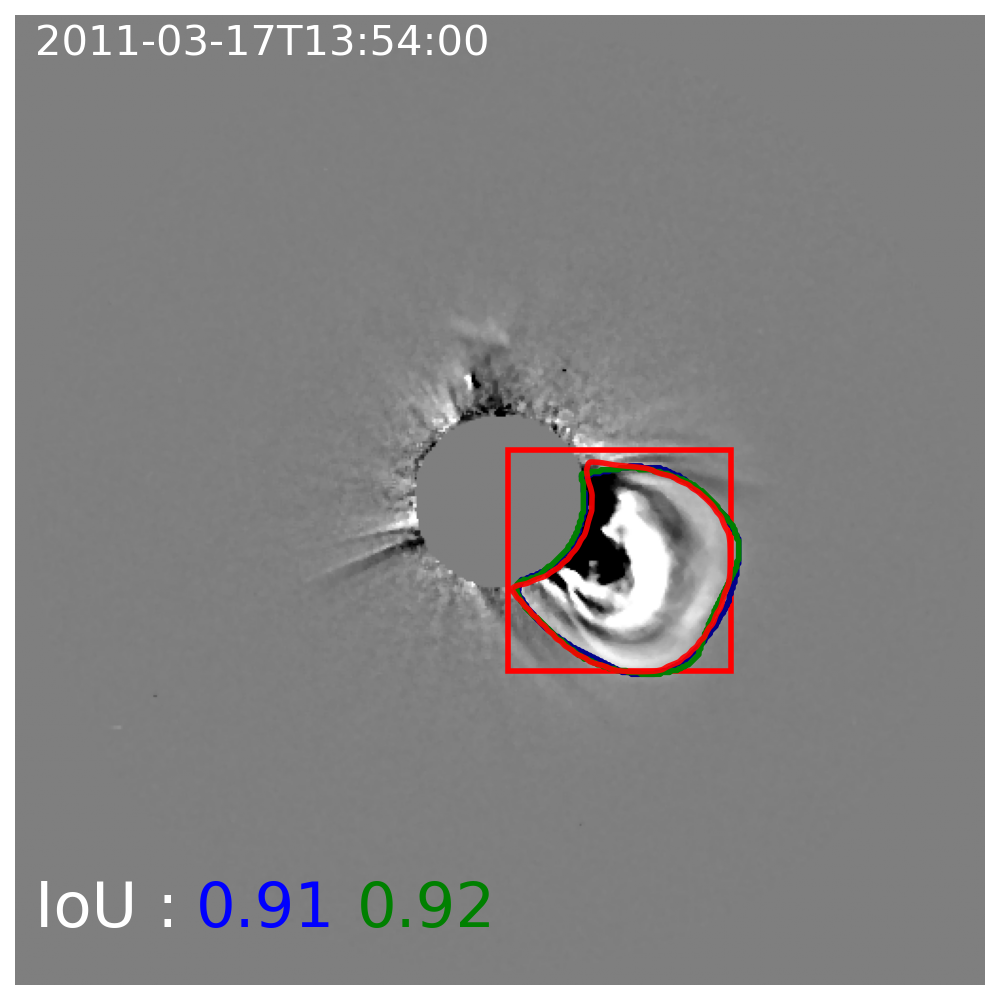}
    \includegraphics[width=0.32\textwidth]{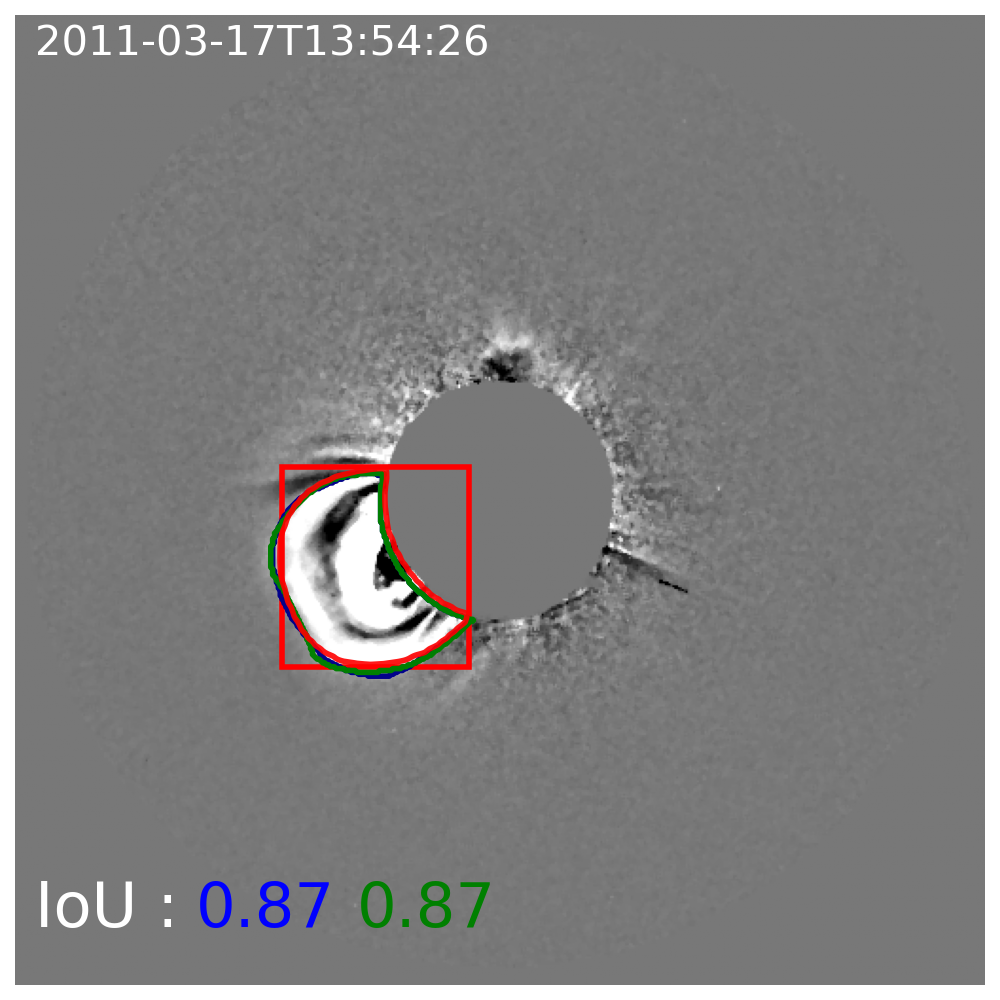}
    \includegraphics[width=0.32\textwidth]{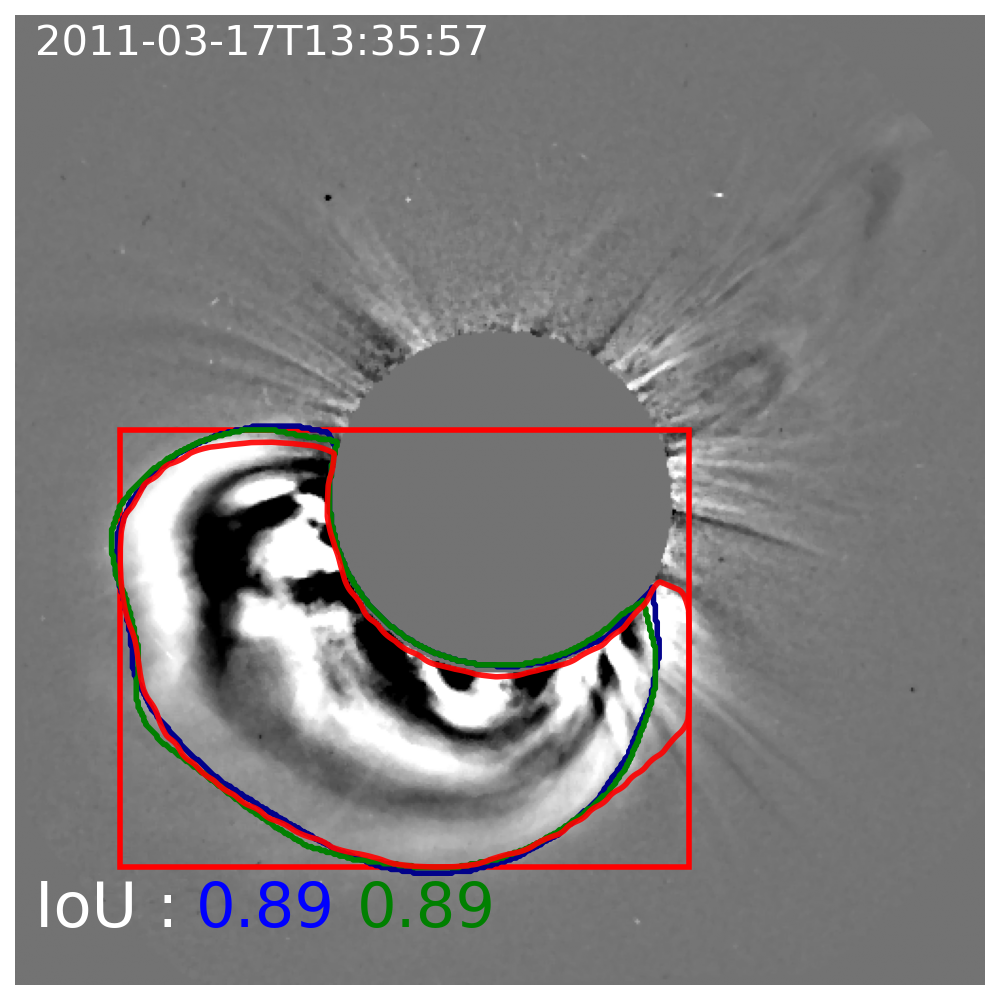}
  \caption{CME outer shell segmentation using our method (\textit{red contour}) versus independent manual estimations by operator 1 (\textit{blue contour}) and 2 (\textit{green contour}). The observations were acquired by COR2-A (\textit{leftmost column}), COR2-B (\textit{middle column}) and LASCO C2 (\textit{rightmost column}). We show cases with $IoU\ge0.75$ between inferred and manual masks, as annotated in the images. See the text for extra details.}
  \label{fig:DNN_masks_on_paper_events}
\end{figure*}

\begin{figure*}
  \centering
    \includegraphics[width=0.32\textwidth]{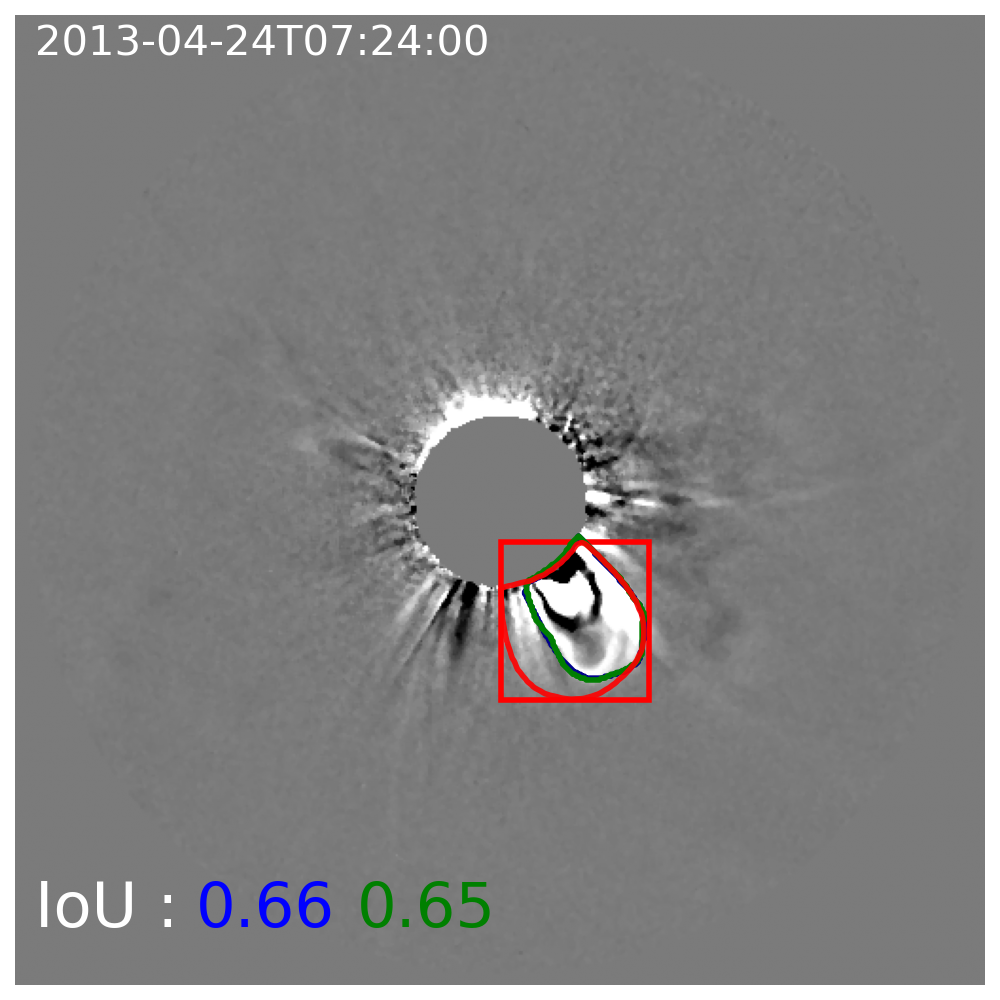}
    \includegraphics[width=0.32\textwidth]{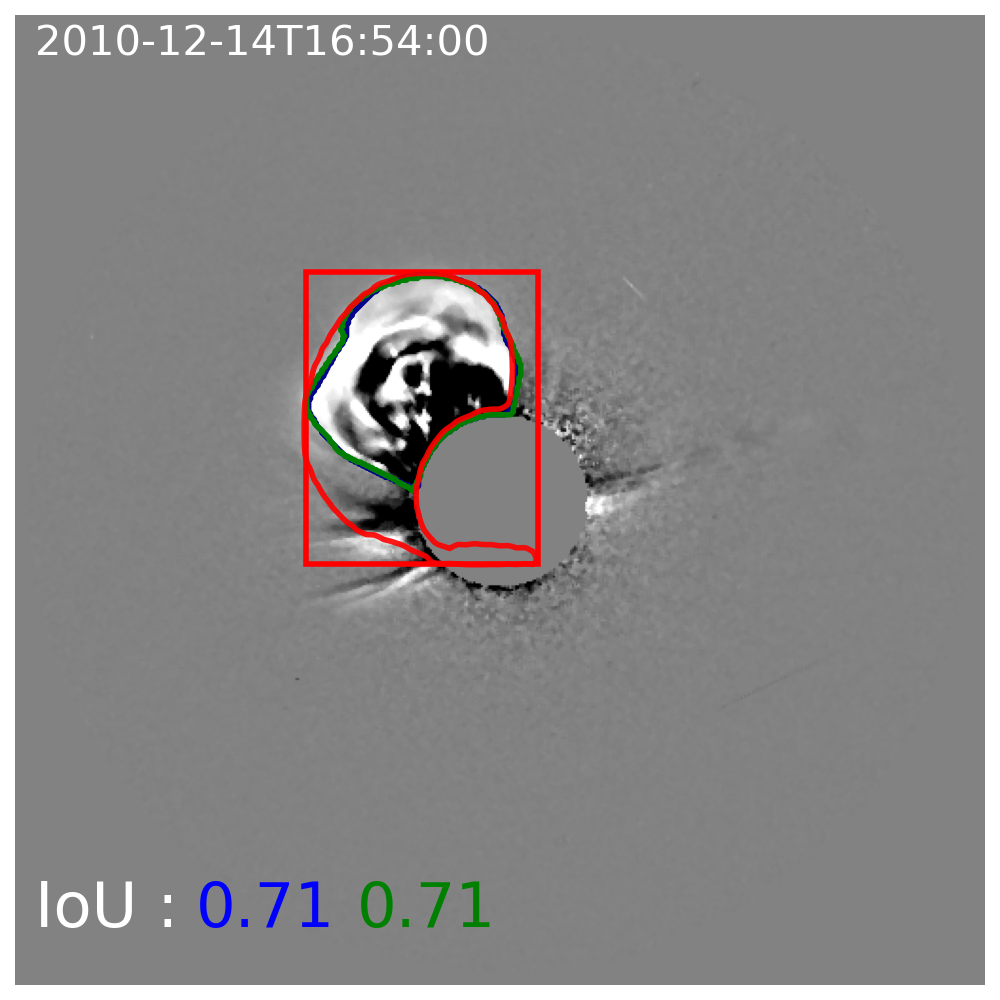}  
    \includegraphics[width=0.32\textwidth]{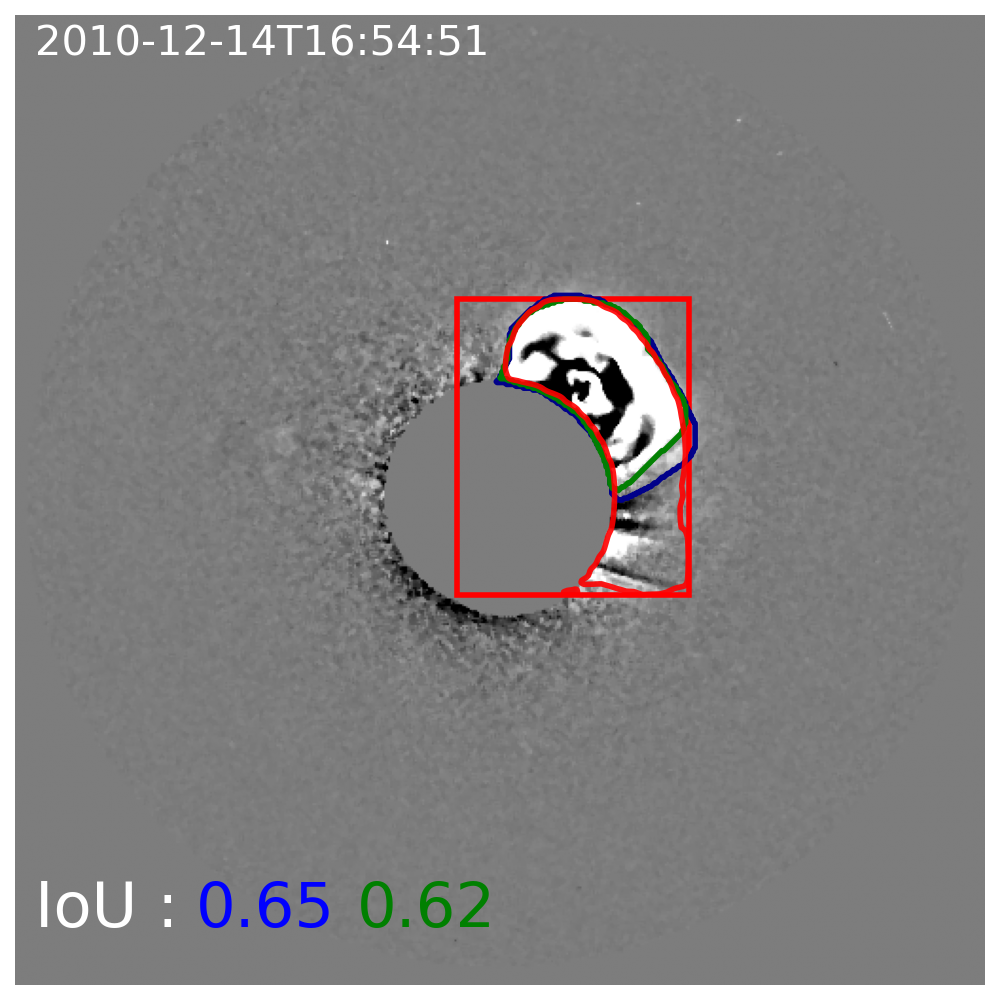}
    \includegraphics[width=0.32\textwidth]{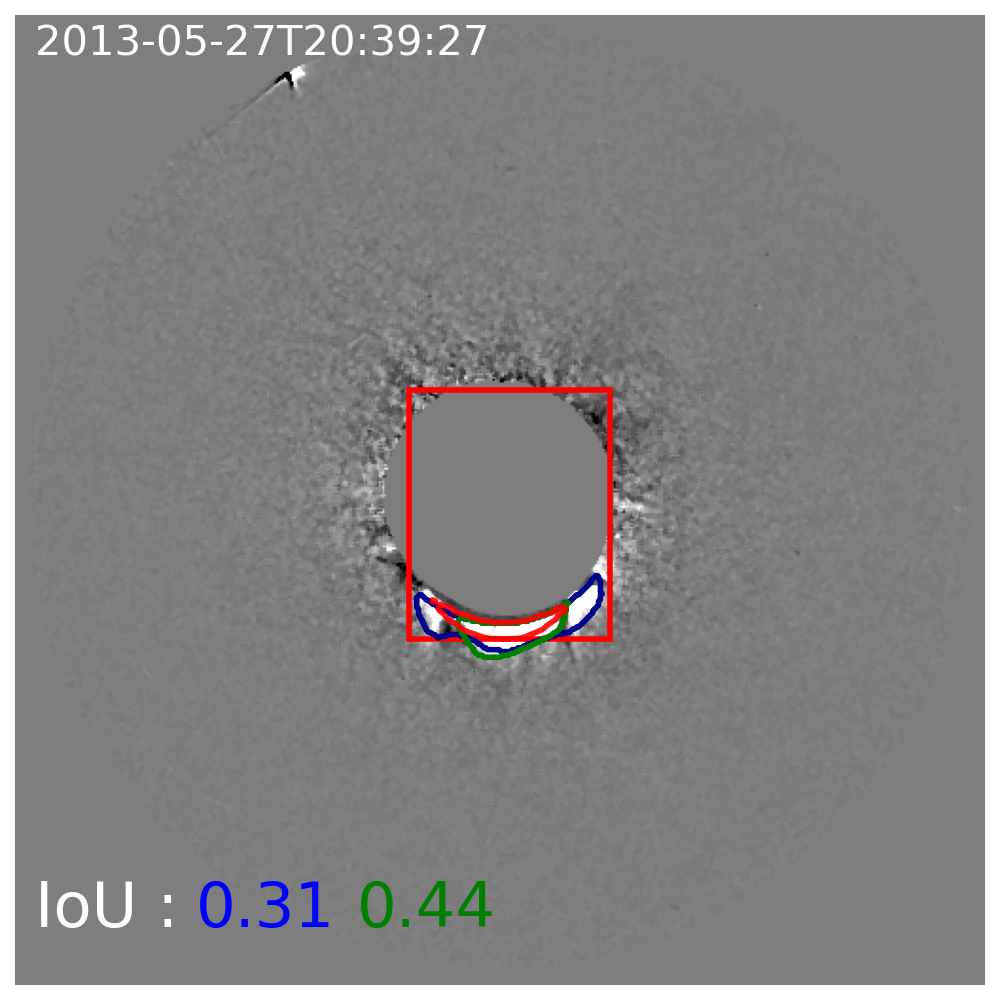}
    \includegraphics[width=0.32\textwidth]{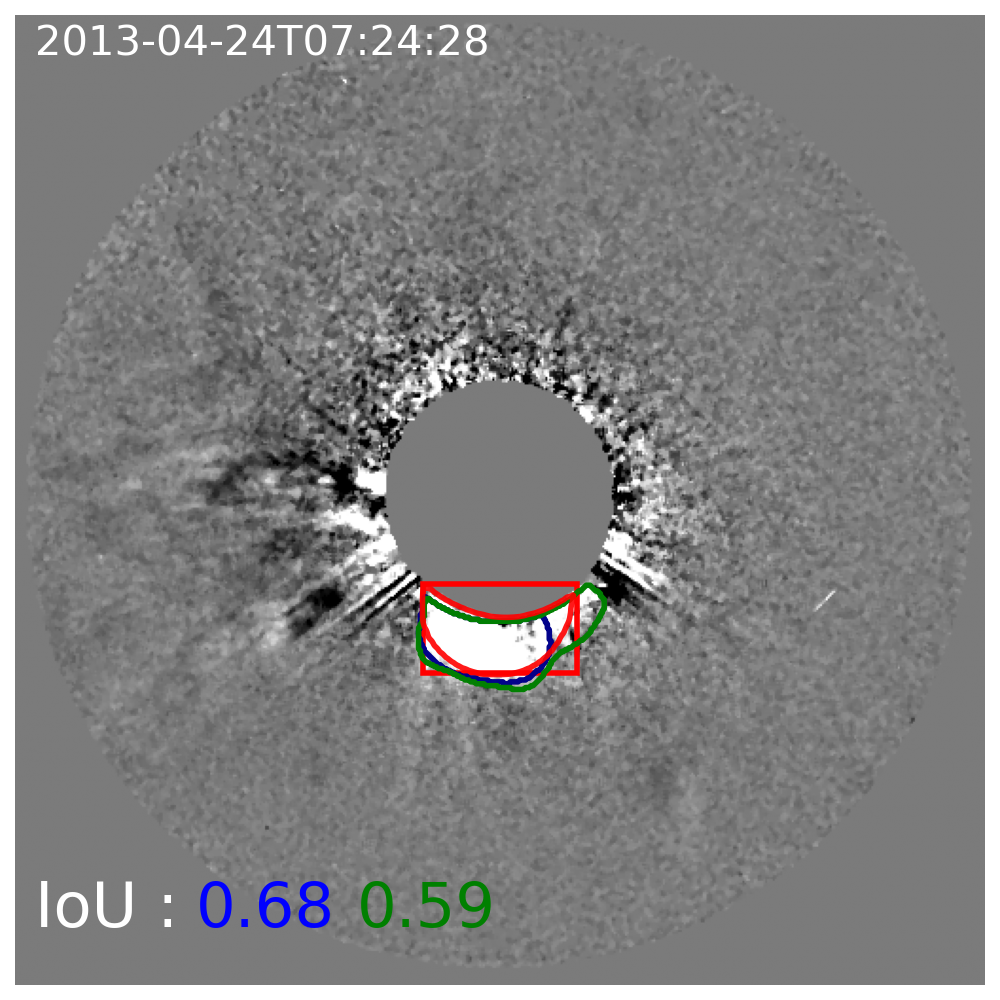}
    \includegraphics[width=0.32\textwidth]{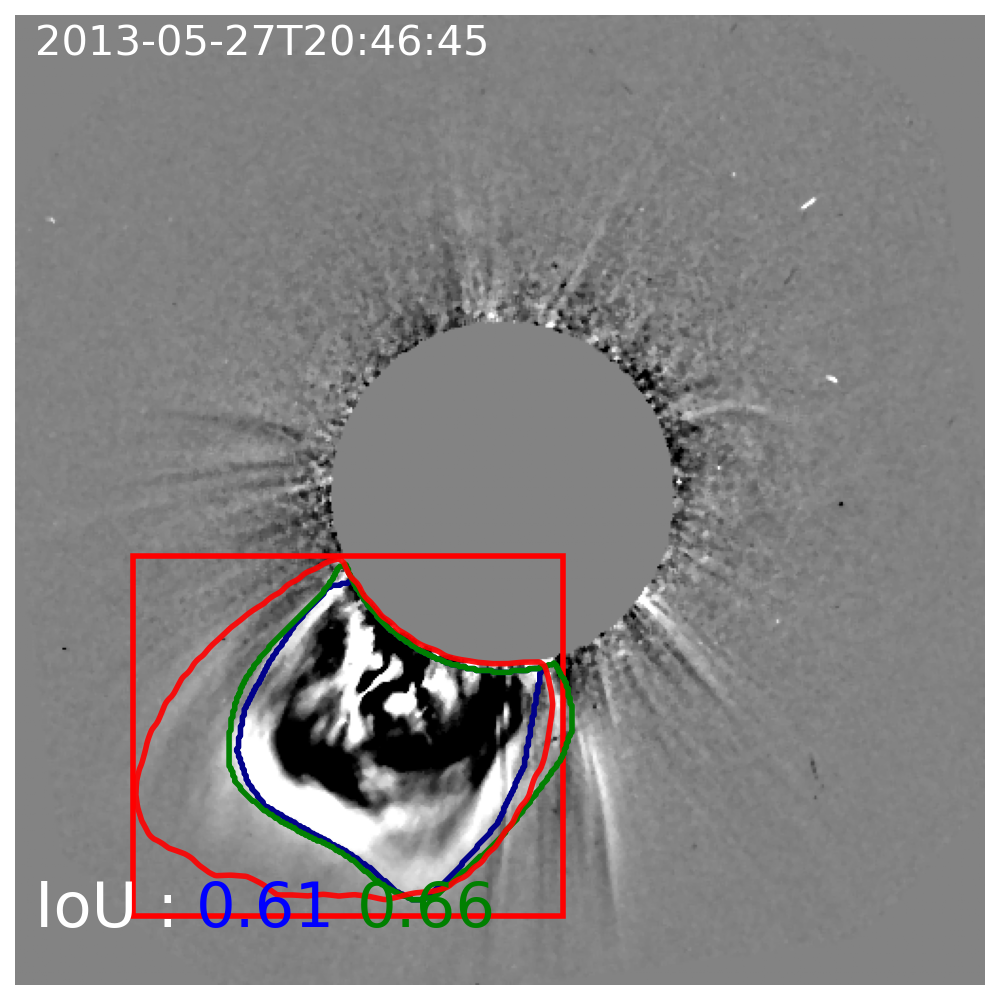}
  \caption{Same as Fig.~\ref{fig:DNN_masks_on_paper_events} but for cases with $IoU<0.75$. The observations were acquired by COR2-A (\textit{panels 1-2}), COR2-B (\textit{panels 3-5}) and LASCO C2 (\textit{last panel}). See the text for extra details.}
  \label{fig:DNN_masks_on_paper_events_bad}
\end{figure*}

To quantify the performance of our instance segmentation, we apply it to 115 images corresponding to the CME events listed in \cite{Cremades2020}, and compare the results with masks derived manually. \textbf{These are images of 12 CME events acquired at different instants and from three different vantage points, by LASCO, COR2-A, and COR2-B. These events have CPA within $25^\circ$ of the solar poles, to minimize projection effects, and have $AWs\ge20^\circ$, therefore, not jet-like or halo CME are included. For more details on the events see \cite{Cremades2020}}. 
The manual segmentations are produced by two independent expert operators by manually selecting points outlying the CME outer envelope. A binary segmentation mask is derived from the selected points using second-order polygons. The histogram of the resulting IoU is shown in Fig.~\ref{fig:res_paper_events_iou_hist}, which indicates that the median $IoU$ is 0.77 ($RSD\approx20-25\%$). As a reference, when comparing the masks of the manual segmentations by operator 1 with operator 2, the median $IoU=0.90$. 
Examples of good ($IoU\ge0.75$) and bad ($IoU\le0.75$) segmentations are given in Fig.~\ref{fig:DNN_masks_on_paper_events} and Fig.~\ref{fig:DNN_masks_on_paper_events_bad}, respectively. We note that in many incorrect segmentations, Mask R-CNN treats nearby structures as part of the CME bulk, an aspect that could be improved by including kinematic information (see Sect.~\ref{sec:conclusions}). Additionally, it is worth noting that in some of the Fig.~\ref{fig:DNN_masks_on_paper_events_bad} cases, there is substantial disagreement even among the independent operators that performed the manual segmentation.

\begin{figure*}
  \centering
  \includegraphics[width=0.49\textwidth]{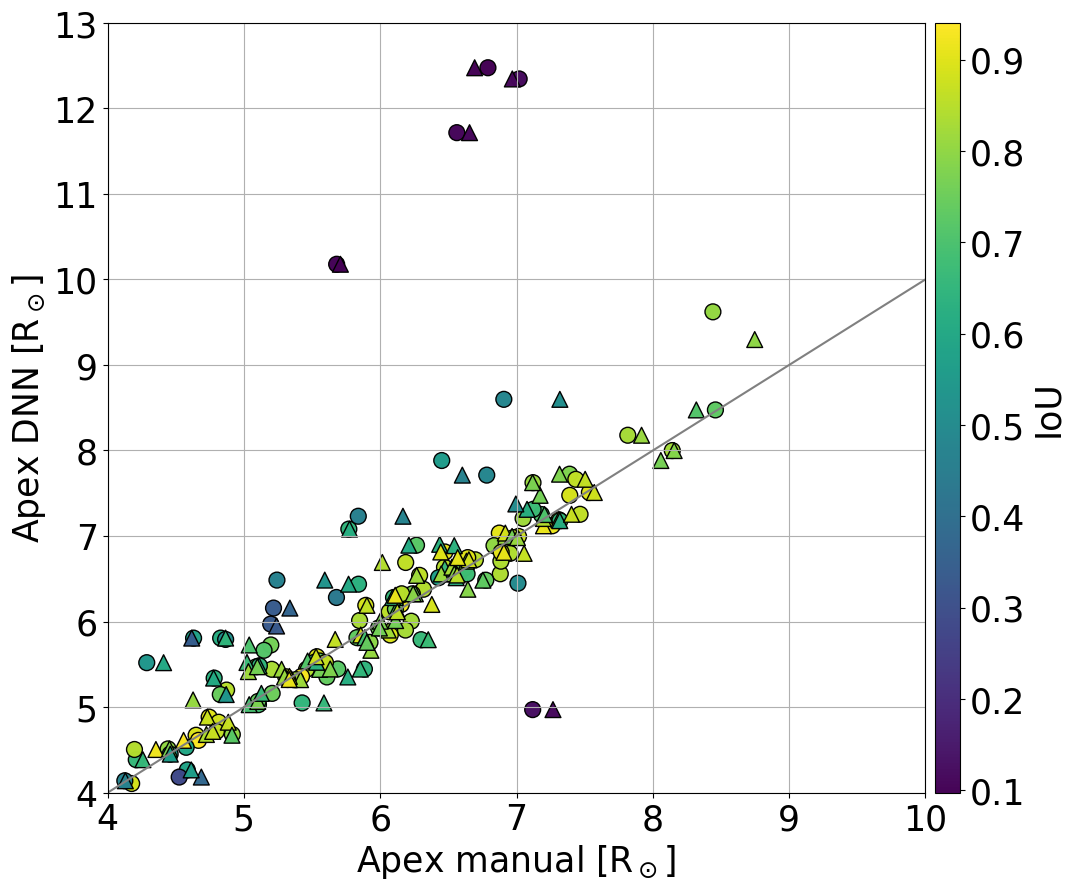}
  \includegraphics[width=0.49\textwidth]{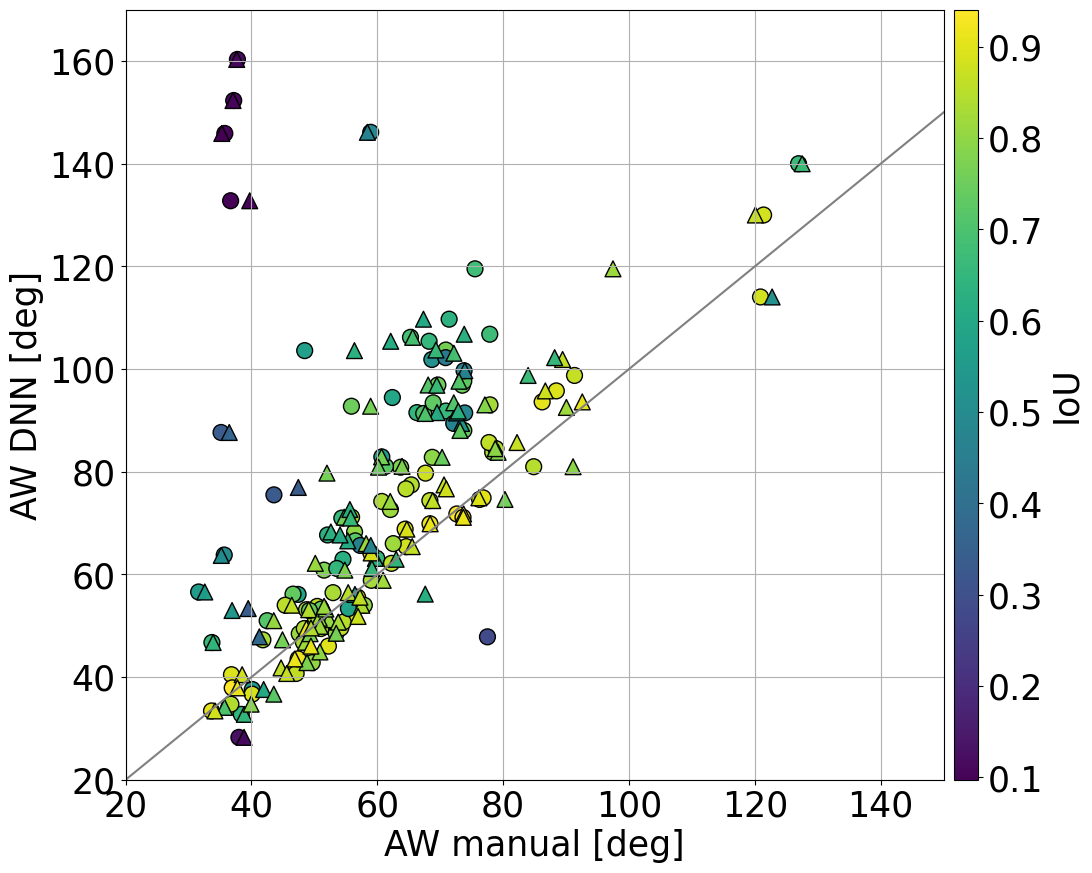}
  \includegraphics[width=0.49\textwidth]{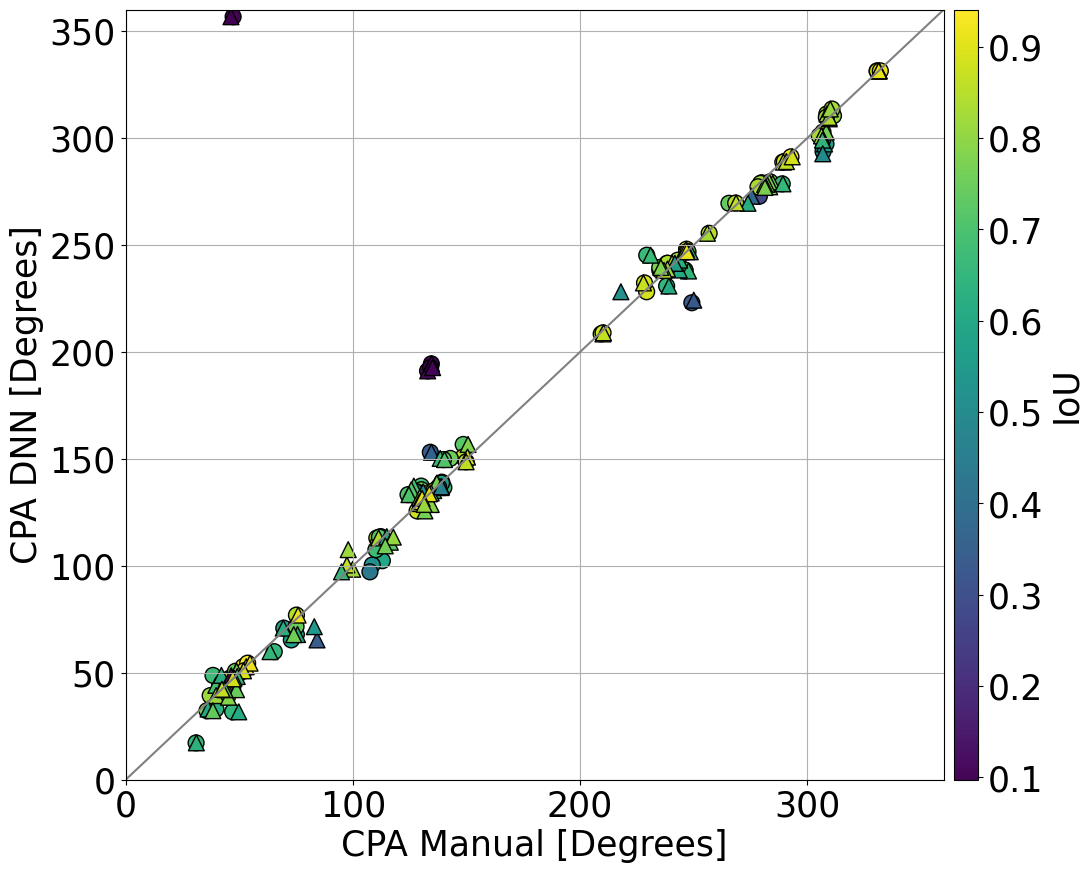}
  \caption{Apex (\textit{top left}), AW (\textit{top right}) and CPA (\textit{bottom}) of the masks obtained with our trained Mask R-CNN (\textit{vertical axes}) and manually (\textit{horizontal axes}). We separate the results of manual operator 1 (\textit{triangles}) and 2 (\textit{squares}). The color bar indicates the $IoU$ score of each image. The solid lines have slope one.}
  \label{fig:res_paper_events_scatter}
\end{figure*}

To link the mask overlapping metrics ($IoU$ and $RSD$) with simpler and commonly used morphological properties, we plot in Fig.~\ref{fig:res_paper_events_scatter} the relation between the apex, AW, and CPA of the manual and inferred masks. Excluding cases with $IoU<0.3$, which are few ($4.8\%$), the best Pearson correlation coefficient is found for CPA ($0.99$), followed by apex ($0.91$) and AW ($0.77$). The latter two are more dependent on the mask shape, thus they can differ even in cases of segmentations with good area overlapping ($IoU\ge0.75$). The overestimation of AW by the DNN in cases with $IoU$ between 0.5 and 0.8 is mostly explained by the fact that the DNN mask may overestimate the CME envelope in regions near the occulter, due to the low signal-to-noise ratio. The overestimation of the apex is more frequent in cases when the outer envelope is faint or when the part of the CME leading shock is wrongly captured.

\begin{figure*}
  \centering
    \includegraphics[width=0.4\textwidth]{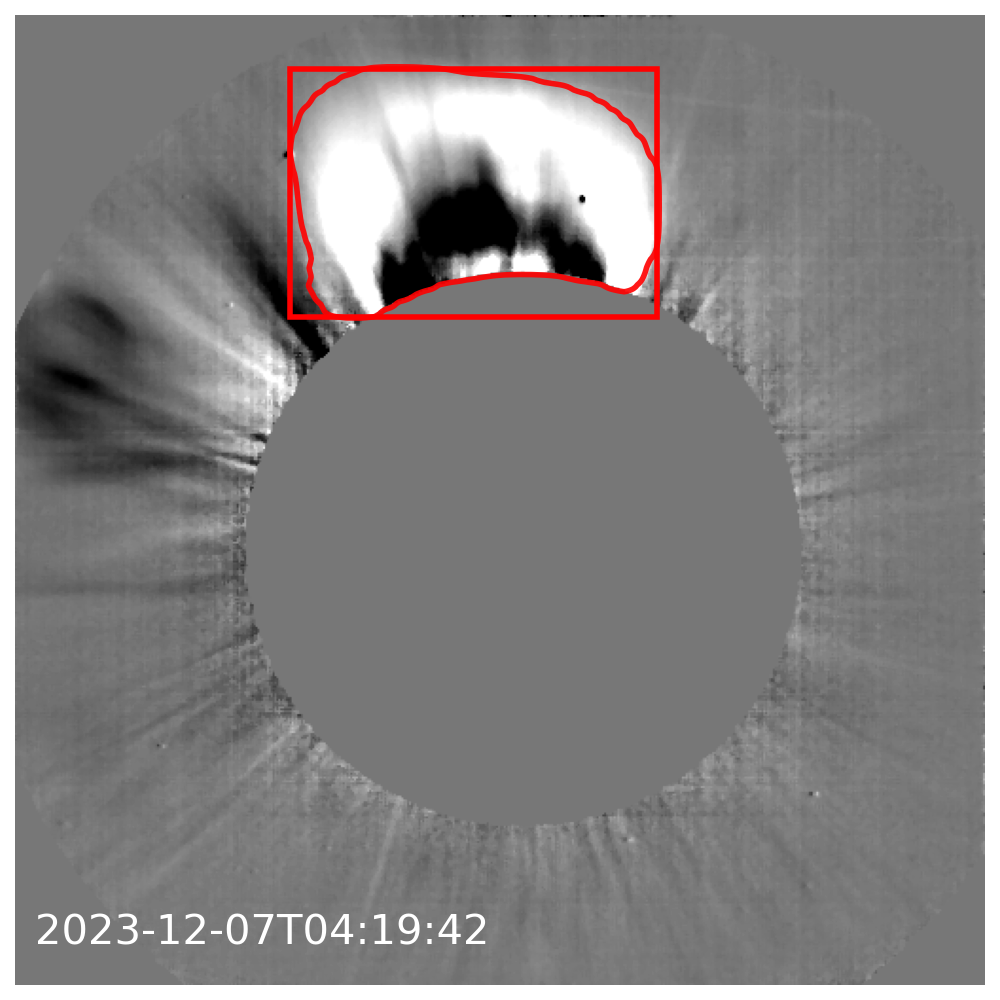}
    \includegraphics[width=0.4\textwidth]{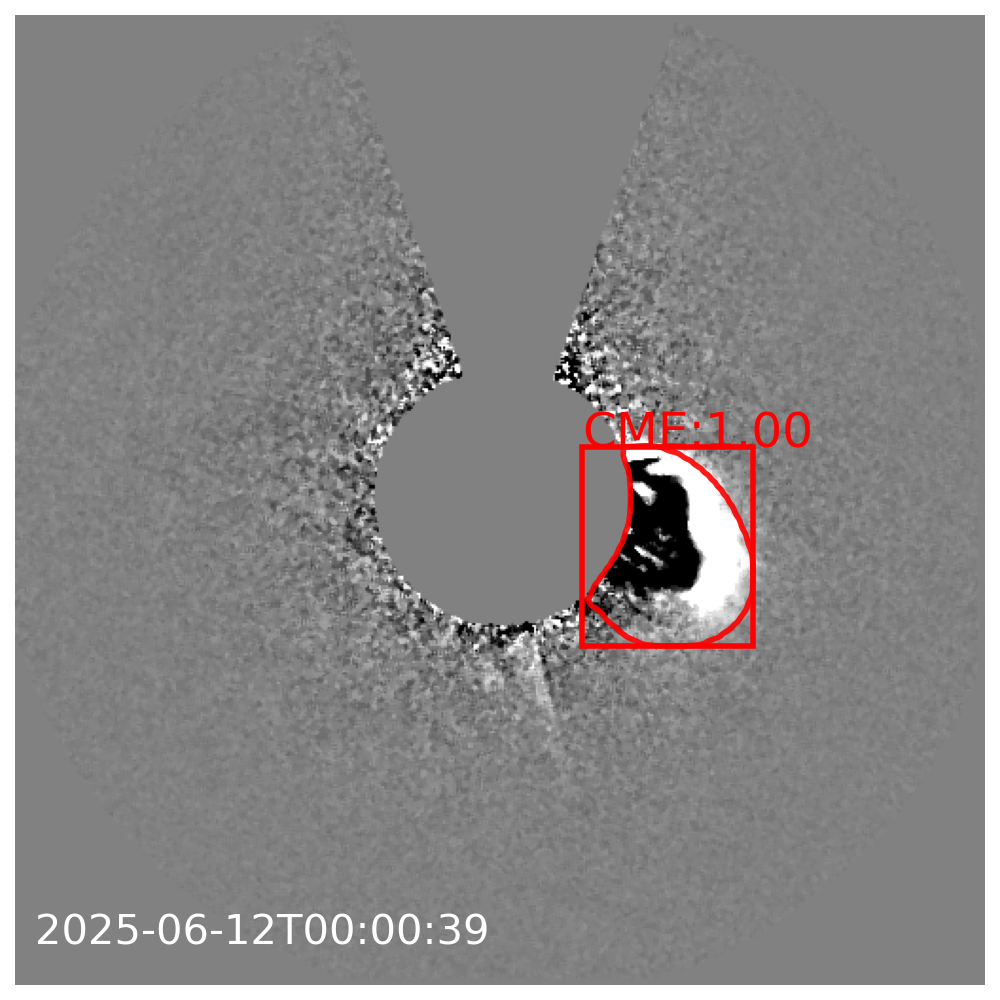}
    \includegraphics[width=0.4\textwidth]{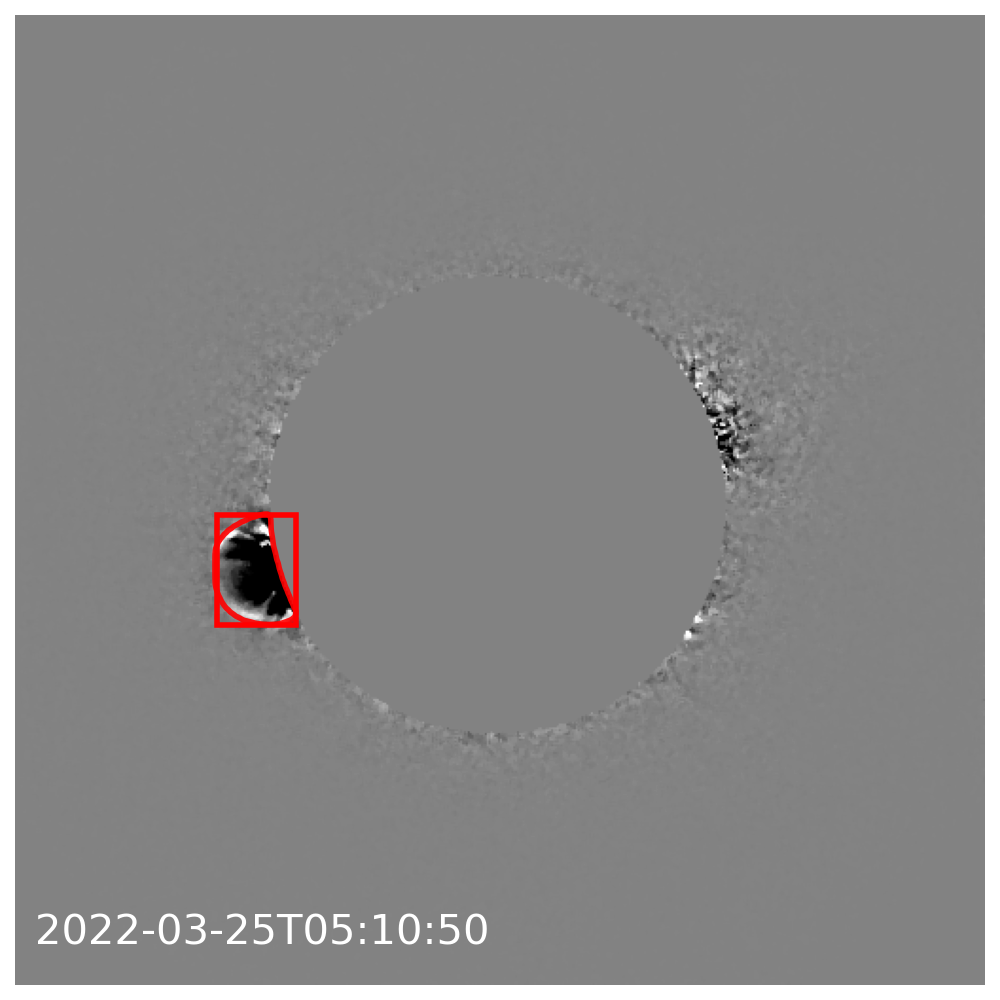}
    \includegraphics[width=0.4\textwidth]{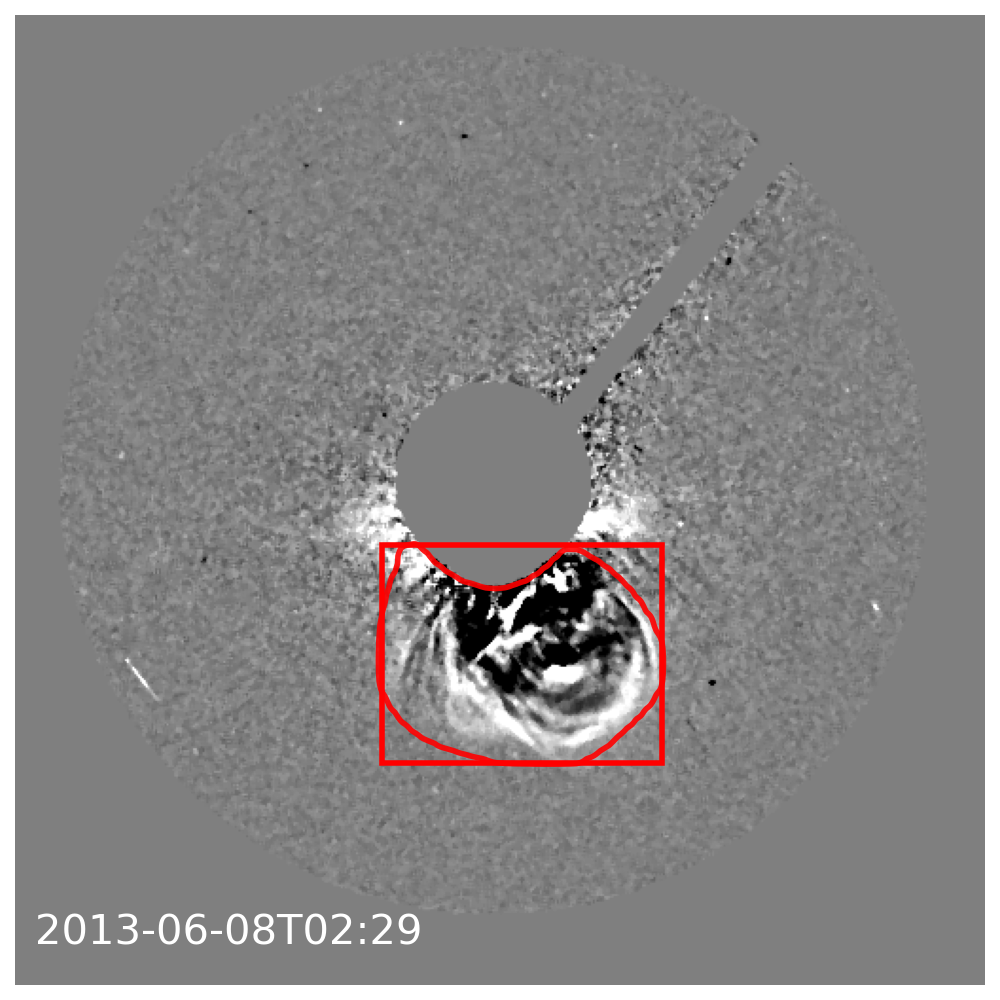}
  \caption{CME segmentation for instruments not included in the Mask R-CNN training dataset. From \textit{left} to \textit{right} and \textit{top} to \textit{bottom}, we show observations by Metis (visible light) onboard Solar Orbiter; by CCOR-1/GOES-19  where we masked out the occulter pylon, by the Full Sun Imager onboard Solar Orbiter at 17.4 nm, where we masked out the solar disk; and by LASCO C3 onboard SoHO.}
  \label{fig:suplementary_DNN}
\end{figure*}

The performance of our method on observations acquired by the instruments included in the training backgrounds (LASCO C2 and COR2-A/B) is demonstrated in Figs.~\ref{fig:obs_mult_cme} to \ref{fig:res_paper_events_scatter}. A significant finding, however, is the generalization capability of our trained Mask R-CNN when applied to observations from other instruments with  distinct fields of view, spatial resolutions, and stray light rejection performance, among others. Even though we do not perform a general quantitative analysis, Fig.~\ref{fig:suplementary_DNN} exemplifies the segmentations obtained for CMEs observed by Metis \citep{antonucci2020, DeLeo2023} onboard Solar Orbiter, by \textbf{the Compact Coronagraphs for the Geostationary Operational Environmental Satellite-19} \citep[CCOR-1/GOES-19,][]{thernisien2025}, by the \textbf{Full Sun Imager} 17.4 nm \citep[Extreme Ultraviolet Imager,][]{rochus2020} onboard Solar Orbiter, and by LASCO C3 \citep{morrill2006} onboard SoHO.

\section{Conclusions and outlook}
\label{sec:conclusions}
We developed a synthetic dataset of CME differential coronagraphic images, and used it to train the commercial-off-the-shelf neural model Mask R-CNN, to perform instance segmentation of the CME outer envelope. Our dataset is generated using an empirical approach that combines coronagraphic observations of the solar corona, with no CME, and synthetic brightness images obtained using a simple CME morphological model, the GCS, and ray-tracing. We have estimated the performance on synthetic validation images, obtaining a median $IoU=0.98$ ($RSD=2\%$), and on manually segmented CME observations acquired by LASCO C2 and COR2-A/B coronagraphs, reaching a median $IoU=0.77$ ($RSD=20-30\%$). 
The model-driven synthetic training allows Mask R-CNN to produce GCS-like (not strictly) masks. These masks are smooth and not representative of the complex exact shape of many CME outer shells. However, they are topologically connected (no holes or isolated patches), which helps to reduce mixing CME bulk material with nearby fast moving structures and segment multiple nearby CMEs, even when no kinematic information is used (other than that given by the single input differential image). Future aspects worth of further investigation include:

\begin{itemize}
    \item We consider that an important improvement of the segmentation can be done by incorporating kinematic information. Namely, by selecting a DNN architecture that can process the whole time series of an event simultaneously, similar to \cite{bauer2025}.
    \item Combine our approach with other neural techniques that can produce more detailed segmentation masks, such as \cite{shan2024} and \cite{lin2024}, to improve their mask filtering step and better reject moving features that are not part of the CME.
    \item We are currently developing a k-means based clustering method to allow the CME tracking procedure (Sect.~\ref{sec:instance_selec}) to detect the temporal sequence of simultaneous multiple CMEs. 
    \item Producing a catalog of segmented CMEs from large observational databases that include faint and narrow events, such as those from SoHO/LASCO and STEREO/CORs, to systematically evaluate performance with respect to manual (CDAW), image processing (CORSET) and, particularly, other neural \citep{shan2024, lin2024} techniques.  
\end{itemize}

Our trained Mask R-CNN also shows an interesting generalization when applied to observations from other instruments with very different properties, including fields of view, spatial resolutions, and stray light rejection performance, among others. This generalization capability is a desired property for any automatic detection tool. Particularly for onboard operation in space platforms, where autonomous identification and prioritization of images is required to exploit the limited telemetry. 

We finally note that the same technique presented in this work can be used with more realistic CME and/or coronal numerical models to produce the synthetic images, while preserving the desired morphological properties of the masks. Notable candidates are 3D magnetohydrodynamic (MHD) CME simulations, such as the Alfvén Wave Solar model \citep[AWSoM,][]{sokolov2013,vanderHolst-etal2014}, that can include flux rope parameters produced by the Eruptive Event Generator based on Gibson-Low \citep[EEGGL,][]{borovikov2017} to initiate the CME simulation; or the coronal 3D MHD model COolfluid COroNal UnsTructured \citep[COCONUT,][]{wang2_2022} that can use a Titov-Démoulin flux rope \citep{titov2014} to initiate the CME simulation. These MHD models could allow defining the CME boundary not only based on brightness spatio-temporal variations, but also on other relevant physical variables, such as density or velocity gradients. 

%
\begin{acks}
Authors acknowledge use of data from the STEREO (NASA) and SOHO (ESA/NASA) missions, produced by the SECCHI and LASCO international consortia. Solar Orbiter is a mission of international cooperation between ESA and NASA, operated by ESA. Solar Orbiter is a space mission of international collaboration between ESA and NASA, operated by ESA. Metis was built and operated with funding from the Italian Space Agency (ASI), under contracts to the National Institute of Astrophysics (INAF) and industrial partners. Metis was built with hardware contributions from Germany (Bundesministerium für Wirtschaft und Energie through DLR), from the Czech Republic (PRODEX) and from ESA. The EUI instrument was built by CSL, IAS, MPS, MSSL/UCL, PMOD/WRC, ROB, LCF/IO with funding from the Belgian Federal Science Policy Office (BELSPO/PRODEX PEA 4000112292); the Centre National d’Etudes Spatiales (CNES); the UK Space Agency (UKSA); the Bundesministerium für Wirtschaft und Energie (BMWi) through the Deutsches Zentrum für Luft- und Raumfahrt (DLR); and the Swiss Space Office (SSO). The Compact Coronagraph-1 (CCOR-1) instrument was designed, built, and tested by the U.S. Naval Research Laboratory (NRL) and is part of NOAA's Space Weather Follow-On (SWFO) Program.
\end{acks}

\begin{fundinginformation}
FAI, HC and FML are members of the ``Carrera del Investigador Cient\'ifico" of CONICET. DGL and FM were supported by CONICET scholarships. FAI, FLC and MST are supported by The Max Planck Partner Group between the University of Mendoza and the Max Planck Institute for Solar System Research (MPS). This research is sponsored by the Department of the Navy, Office of Naval Research Global, under Grant Award number N629092512017. Any opinions, findings, and conclusions or recommendations expressed in this material are those of the authors and do not necessarily reflect the views of the Office of Naval Research. This work is also part of the DynaSun project and has thus received funding under the Horizon Europe programme of the European Union under grant agreement no. 101131534. Views and opinions expressed are however those of the authors only and do not necessarily reflect those of the European Union and therefore the European Union cannot be held responsible for them.

\end{fundinginformation}

\begin{dataavailability}
\end{dataavailability}
%
%
 \bibliographystyle{spr-mp-sola}
\bibliography{iglesias}  
%
%
%
%

\end{article} 
\end{document}